\documentclass[final,5p,times]{elsarticle}

\usepackage{amssymb}

\usepackage{amsmath,amssymb,amsfonts}
\usepackage{algorithm,algorithmic}
\usepackage{graphicx}
\usepackage{textcomp}
\usepackage{gensymb}
\usepackage{hyperref}
\usepackage[disable]{todonotes} 
\usepackage{multirow}
\usepackage{booktabs}
\usepackage{tabularx}
\usepackage{placeins}
\usepackage{mathrsfs}
\usepackage{float}
\usepackage{caption}
\usepackage{subcaption}
\usepackage{array}
\usepackage{makecell}

\usepackage{enumitem}
\usepackage{bm} 
\usepackage{tikz}
\usepackage{pgfplots}
\pgfplotsset{compat=1.18}
\usepackage{pgfplotstable,booktabs}
\usetikzlibrary{pgfplots.groupplots}
\usepgfplotslibrary{dateplot,fillbetween}
\usepackage{pgf}
\usepgfplotslibrary{colorbrewer}
 \usepackage[nolist]{acronym} 
 
\DeclareMathOperator{\R}{\mathbb{R}}

\newcommand{\changeA}[1]{\textcolor{black}{#1}}
\newcommand{\change}[1]{\textcolor{black}{#1}}

\hypersetup{pdfauthor={Name}} 

\graphicspath{{figures/}}

\journal{Elsevier}

\begin{document}

\begin{frontmatter}

\title{\change{User-specified} Kelvin-hour budgets within model predictive control for energy-efficient buildings: Simulation and field demonstration}

\author[label1]{Jicheng Shi\corref{cor1}}
\ead{jicheng.shi@epfl.ch}
\author[label1]{Colin N. Jones}

\cortext[cor1]{ Corresponding author}

\affiliation[label1]{organization={Automatic Control Lab},
            addressline={EPFL}, 
            city={Lausanne},
            postcode={1015}, 
            state={Vaud},
            country={Switzerland}}

\begin{acronym} 
\acro{mpc}[MPC]{model predictive control}
\acro{kh}[Kh]{Kelvin-hour}
\acro{hvac}[HVAC]{heating, ventilation and air-conditioning}
\acro{hp}[HP]{heat pump}
\acro{ghi}[GHI]{global horizontal irradiance}
\acro{hdd}[HDD]{heating degree-days}
\acro{arx}[ARX]{auto-regressive with exogenous input}
\acro{rc}[RC]{resistance-capacitance}
\acro{deepc}[DeePC]{data-enabled predictive control}
\acro{qp}[QP]{quadratic programming}
\acro{nlp}[NLP]{nonlinear programming}
\acro{cop}[COP]{coefficient of performance}
\acro{boptest}[BOPTEST]{Building Optimization Testing Framework}
\acro{scp}[SCP]{split conformal prediction}
\acro{pmv}[PMV]{predicted mean vote }
\acro{ppd}[PPD]{predicted percentage of dissatisfied}
\acro{daddpc}[DAD-DPC]{disturbance-adaptive data-driven predictive control}
\acro{iid}[i.i.d.]{independent and identically distributed}
\acro{prbs}[PRBS]{pseudo-random binary signal}
\acro{nnarx}[NNARX]{neural-network-based nonlinear auto-regressive with exogenous input}
\acro{lti}[LTI]{linear time-invariant}
\end{acronym}

\begin{abstract}
For energy-efficient building control, \acf{mpc} \change{has been widely proposed} to reduce energy use while maintaining prescribed indoor temperature bounds. In practice, \acs{mpc} implementations may still produce temperature-bound violations because of model mismatch, weather-forecast errors, and softened constraints introduced to preserve feasibility. Existing soft-constrained \acs{mpc} usually handles this trade-off through slack-weight tuning, so \change{users} cannot prescribe a violation-severity budget before operation. This paper develops a Kelvin-hour (\acs{kh})-budget \acs{mpc} that \change{allows users to specify a} running-average budget for temperature-bound violation severity. 
During operation, the controller uses the realized \acs{kh} budget surplus or overspend to update the temperature bounds, providing step-by-step budget feedback.
We evaluate the method in two high-fidelity \acs{boptest} simulation cases and in an occupied residential deployment.
In the first one-zone case, prescribed budgets produce clear running-average \acs{kh} violation responses across predictors and disturbance settings. At \change{a prescribed budget of} $0.005~\mathrm{Kh/step}$, the controller reduces energy use by $31.9\%$ and \acs{kh} violation by $36.2\%$ relative to the built-in default controller.
In the second coupled two-zone case, separate zone budgets produce zone-level responses, with $18.9\%$--$20.5\%$ energy reduction and  $9.4\%$--$64.3\%$ zone-level Kh violation reduction. In the occupied residential deployment, the running-average \acs{kh} violations remain close to or below their prescribed budgets under real sensing, actuation, weather, and occupancy conditions.
\end{abstract}




\end{frontmatter}

\section{Introduction}
\label{sec:intro}

Building climate control aims to reduce energy use while maintaining indoor thermal comfort under changing weather, occupancy, and equipment constraints. \Acf{mpc} is well suited to this problem because it can coordinate building thermal inertia, comfort bounds, actuator limits, and energy objectives in a receding-horizon optimization~\cite{drgovna2020all,killian2016ten}. In most \acs{mpc} formulations for buildings, thermal comfort is represented by temperature bounds~\cite{drgovna2020all}. However, during online operation,  temperature-bound violations can still occur due to plant-model mismatch, imperfect weather forecasts, and actuator limits~\cite{sturzenegger2015model,hou2022model}.

Such temperature-bound violations are usually managed indirectly through slack penalties for soft constraints~\cite{drgovna2020all,khabbazi2025,huang2024overlooked}, or through chance levels, scenario choices, and constraint-tightening margins in uncertainty-aware formulations~\cite{oldewurtel2012use,parisio2014implementation,yin2024data}. These quantities are meaningful \change{for controller design and tuning}, but they leave \change{a practical question}: before operation begins, how can \change{the user specify} the allowed severity of temperature-bound violations \change{for} closed-loop operation? \change{Here, the user is the person or team responsible for setting such operating limits, such as a resident in a home or a building operations team in a larger building.} This work addresses this challenge with a \acf{kh}-budget \acs{mpc}, where the \change{user} pre-specifies the allowed severity as a running-average \acs{kh} budget. The controller uses this budget for step-by-step feedback, and we assess the empirical budget response across predictor classes, measurement-noise levels, weather-forecast-error groups, coupled multi-zone operation, and an occupied residential deployment.

\subsection{Literature review on comfort-violation control}
\label{sec:intro_lit}

Comfort-bound violations can be summarized with different operational meanings. Common metrics include violation-rate measures, \acs{kh} violation, and comfort-model-based indices such as \acf{pmv} and \acf{ppd}~\cite{carlucci2012review,carlucci2013book}. These metrics are also commonly reported in evaluations of building \acs{mpc} performance~\cite{huang2021simulation,zanetti2023performance,yang2020model,shi2025adaptive}.
The budget problem considered in this work concerns the severity of temperature-bound violations during operation.
Among metrics based directly on bound crossings, the violation rate indicates how often the temperature band is crossed, but it does not distinguish a shallow excursion from a deep one. By contrast, \acs{kh} combines how far and how long the temperature remains outside the band. Comfort-model-based indices remain useful for comfort assessment, but require additional assumptions about the thermal environment and occupant-related parameters~\cite{carlucci2012review,ashrae2023thermal}.
Therefore, \acs{kh} is used as the budget variable in this paper. It can be computed directly from temperature-bound exceedances and updated during online operation.

Softened comfort constraints are common in \acs{mpc} for building systems, as strict temperature-bound satisfaction may be infeasible due to prediction errors or actuator limits~\cite{sturzenegger2015model,khabbazi2025}.
A non-negative slack variable relaxes the predicted temperature bounds, while its penalty in the objective sets how strongly predicted violations are avoided relative to energy use or operating cost.
This design is commonly evaluated by reporting energy use together with temperature-bound discomfort in \acs{kh}~\cite{zanetti2023performance,stoffel2023evaluation,arroyo2022comparison,freund2021implementation}. For example, in a one-year closed-loop simulation, white-box, gray-box, and black-box \acs{mpc} variants saved 4.9\%--8.4\% energy and reduced annual \acs{kh} discomfort by 7.8\%--83.8\% relative to a well-tuned rule-based controller~\cite{stoffel2023evaluation}.
Slack-weight choices can strongly affect the achieved energy-discomfort outcome~\cite{huang2024overlooked,zheng2024economic}. In a residential \acs{mpc} simulation study, a well-tuned slack weight setting achieved 17\%--34\% energy cost savings and 30\%--95\% discomfort reduction relative to rule-based control~\cite{zheng2024economic}. However, the full weight sweep produced cost-discomfort curves that changed with the controller model and testing season.
Thus, the slack weight can shape the realized \acs{kh} discomfort, but it does not specify a pre-operation \acs{kh} budget.

Stochastic \acs{mpc} makes uncertainty explicit in the temperature bounds~\cite{mesbah2016stochastic}. These formulations use chance constraints, scenario sets, disturbance distributions, or forecast-error models to limit temperature-bound violations under uncertainty~\cite{oldewurtel2012use,parisio2014implementation,uytterhoeven2022chance,mohebi2025chance}. 
For example, a recent data-driven stochastic \acs{mpc} study accounts for measurement noise and weather-forecast errors, reporting up to 90\% improvement in constraint satisfaction and 8\% energy savings~\cite{yin2024data}.
A related direction constrains the running-average violation rate over time~\cite{korda2012stochastic,oldewurtel2013adaptively,shi2025dad,shi2024disturbance}. By using past violations to adjust chance constraints at the current step, this formulation can reduce conservatism~\cite{korda2012stochastic}. 
Such running-average constraints also match comfort-evaluation practice, where discomfort is often expressed as averaged or accumulated indices over an operating period~\cite{carlucci2012review}.
More recently, \acf{daddpc} tested this idea with disturbance adaptation in a real-world campus building, regulating a 5\% average violation rate with 20.5\% energy savings relative to the default controller~\cite{shi2024disturbance}.
These stochastic and running-average methods make the violation rate more explicit and controllable, and several also report \acs{kh} violation after operation~\cite{oldewurtel2012use,yin2024data,mohebi2025chance,shi2024disturbance}. However, the target quantity is the violation rate, not a prescribed \acs{kh} budget for violation severity.

The closest precedents for a pre-specified \acs{kh} budget are severity-aware violation   formulations. 
A control-theoretic precedent is the stochastic \acs{mpc} framework in~\cite{korda2014stochastic}, which regulates violation severity through a general severity function. Its building control example quantifies temperature-bound violation in \acs{kh}.
Related iterative-learning stochastic \acs{mpc} targets linear building \ac{hvac} systems with adaptive constraint tightening~\cite{long2020iterative}.
The tightening is updated iteratively so that the empirical average of a violation-severity measure converges to a prescribed expectation in probability.
A separate data-driven approach uses Bayesian optimization to tune rule-based building controllers subject to daily cumulative-discomfort constraints expressed in \acs{kh}~\cite{xu2024pdcbo}.
These studies show that violation severity can be specified in control or tuning problems. However, for deployment,  two gaps remain:
\begin{itemize}[nolistsep,leftmargin=1.5em]
    \item \textbf{\textit{Operational feedback:}} The iteration-level constraint adaptation in~\cite{long2020iterative} and the daily formulation in~\cite{xu2024pdcbo} do not provide step-by-step budget feedback on the realized \acs{kh} account. However, occupants experience discomfort during operation rather than only through end-of-iteration or end-of-day summaries. 
    \item \textbf{\textit{Deployment assumptions:}} The methods in~\cite{korda2014stochastic,long2020iterative} are designed and analyzed for \acf{lti} systems subject to \acf{iid} disturbances.
     These assumptions are difficult to maintain in building operation, where building and \acs{hvac} dynamics can include nonlinear, time-varying, and coupled processes~\cite{afroz2018modeling}, and where forecast errors, unmeasured gains, and occupancy effects often lead to non-\acs{iid} disturbance sequences~\cite{gholamzadehmir2020adaptive,zheng2025quantifying}.
\end{itemize}
These gaps are critical because they determine whether a \change{prescribed} \acs{kh} budget can \change{be used} in real building control. \change{A deployable approach} needs step-by-step feedback during operation and should not rely on assumptions that are difficult to establish in real buildings.

We address these two gaps with a \acs{kh}-budget \acs{mpc}. First, to provide step-by-step budget feedback, the controller uses signed bound adaptation at each step. The realized \acs{kh} violation is compared with the prescribed budget: a budget surplus relaxes the temperature bounds for energy reduction, whereas a budget overspend reduces this relaxation. 
Second, the implementation and empirical evaluation are not tied to an \acs{lti} predictor or \acs{iid} disturbances. We evaluate the proposed \acs{kh}-budget \acs{mpc} across fixed and online-adaptive predictors, linear and nonlinear models, measurement-noise levels, correlated weather-forecast-error conditions, coupled two-zone operation, and an occupied residential deployment.
This use of realized violations to adjust future temperature bounds is related to the \ac{daddpc} framework for violation-rate regulation~\cite{shi2024disturbance}.
However, in the present work, the feedback is based on \acs{kh} budget surplus or overspend rather than violation rate, so the \change{user} specifies violation severity. The proposed controller also removes the \ac{daddpc} requirement for a separately designed conservative backup controller and uses signed bound adaptation throughout operation.

\subsection{Contributions and paper structure}
\label{sec:intro_contrib}

The paper makes three contributions:
\begin{enumerate}[]
\item We propose a \acs{kh}-budget \acs{mpc} that formulates comfort management \change{through a user-specified running-average} \acs{kh} violation-severity budget. The controller converts the \acs{kh} budget surplus or overspend into step-by-step signed bound adaptation.
\item We validate the proposed controller in two high-fidelity simulation cases. 
In the first case, prescribed \acs{kh} budgets produce clear running-average responses across predictors and disturbance settings. At $\alpha=0.005~\mathrm{Kh/step}$, the controller reduces energy use by $31.9\%$ and the \acs{kh} violation by $36.2\%$ relative to the built-in default controller. 
In the second coupled two-zone case, separate zone budgets produce zone-level responses, with $18.9\%$--$20.5\%$ energy reduction and $9.4\%$--$64.3\%$ zone-level \acs{kh}-violation reduction.
\item We implement the proposed controller in an occupied residential home with a commercial heat pump. Across three six-day field tests, the running-average \acs{kh} responses are close to or below the prescribed budgets under real sensing, actuation, weather, and occupancy conditions.
\end{enumerate}

The remainder of this paper is organized as follows. Section~\ref{sec:method} defines the preliminaries for energy-efficient building \acs{mpc} and presents the \acs{kh}-budget MPC algorithm. Section~\ref{sec:case_studies} introduces the simulation and field configurations. Sections~\ref{sec:results_sim} and~\ref{sec:results_exp} report the simulation and occupied-home results, respectively. Section~\ref{sec:conclusion} presents the conclusions and discusses the limitations and future work.

\section{Methodology}
\label{sec:method}

This section introduces the modeling setup, the nominal \acs{mpc}, and the proposed \acs{kh}-budget \acs{mpc} framework. Section~\ref{sec:method_prelim} defines the building model, temperature bounds, and \acs{kh} violation. Section~\ref{sec:method_nominal} presents the nominal \acs{mpc} with soft constraints, and Section~\ref{sec:method_budget} introduces the proposed \acs{kh}-budget \acs{mpc}.

\subsection{Preliminaries}
\label{sec:method_prelim}

\textbf{\textit{Building model}} Consider a building zone observed at discrete time steps $t=0,1,\ldots$ with sampling interval $dT$ expressed in hours per step. The state, control input, and measurable disturbance are denoted by $x(t)\in\R^{n_x}$, $u(t)\in\R^{n_u}$, and $w(t)\in\R^{n_w}$, respectively. The disturbance vector primarily contains outdoor temperature and \acf{ghi}. The building thermal dynamics can be modeled using different methods, including \acf{rc} models, \acf{arx}, and neural networks~\cite{drgovna2020all}. We write the model abstractly as
\begin{equation} \label{eq:f}
  x(t+1) = f\!\left(x(t),u(t),w(t)\right).
\end{equation}
The comfort-relevant output $y(t)\in\R$, such as zone operative temperature or representative zone-air temperature, is obtained through a known map
\begin{equation*}
  y(t) = g\!\left(x(t)\right).
\end{equation*}
For example, if $x(t)$ contains the room, wall, and floor temperatures in an \acs{rc} model~\cite{zanetti2023performance}, then $g\!\left(x(t)\right) = [1\ 0\ 0]x(t)$ selects the room temperature as $y(t)$.

\textbf{\textit{Temperature bounds}} The nominal comfort requirement is represented by a time-varying temperature interval,
\begin{equation*}
  T^l(t) \le y(t) \le T^u(t),
\end{equation*}
where the lower and upper bounds may change with occupancy schedules. 

\textbf{\textit{\acs{kh} violation}} The corresponding per-step temperature violation magnitude is
\begin{equation}
  \hat{v}(t) =
  \max\!\left(0,\; T^l(t)-y(t),\; y(t)-T^u(t)\right),
  \label{eq:vtT}
\end{equation}
and the per-step \acs{kh} violation expressed in $\mathrm{Kh/step}$ is
\begin{equation}
  v(t) = \hat{v}(t) \, dT.
  \label{eq:vt}
\end{equation}

\subsection{Nominal \acs{mpc} with soft constraints for energy-efficient building control}
\label{sec:method_nominal}

In a receding-horizon implementation, the nominal MPC with soft constraints solves the following optimization problem at each time step $t$ and applies only the first input of the optimal sequence:
\begin{subequations}
\label{eq:nominal_mpc}
\begin{align}
  \min_{u_{i|t},\,s_{i|t}} & \quad
   \sum_{i=0}^{N-1}
  l_{i|t}\!\left(x_{i|t},u_{i|t},w_{i|t}\right)
  + \sum_{i=1}^{N}\rho_\mathrm{slack} \, s_{i|t}^2
  \label{eq:nominal_mpc_obj} \\
  \mathrm{s.t.}\quad
  & x_{0|t} = x(t), \\
  & x_{i+1|t} =
  f\!\left(x_{i|t},u_{i|t},w_{i|t}\right),  \, i=0, \ldots,N-1 \\
  & u_{i|t} \in \mathbb{U}, \quad i=0,\ldots,N-1,\\
  & y_{i|t} =
  g\!\left(x_{i|t}\right),
  \ i=1,\ldots,N, \\
  & T^l_{i|t} - s_{i|t}
  \le y_{i|t} \le
  T^u_{i|t} + s_{i|t}, 
   \ i=1, \ldots,N, \label{eq:nominal_mpc_slack}\\
  & s_{i|t} \ge 0,
  \ i=1, \ldots, N.
\end{align}
\end{subequations}
In this notation, $x(t)$ is the current state at time $t$, while $x_{i|t}$ is the state predicted $i$ steps ahead using information available at time $t$. The same convention is used for the predicted output $y_{i|t}$, the predicted input $u_{i|t}$, the weather forecast $w_{i|t}$, the temperature bounds $T^l_{i|t}$ and $T^u_{i|t}$, and the slack variable $s_{i|t}$. \change{The set $\mathbb{U}$ defines the admissible control inputs, including actuator limits.} The term \change{$l_{i|t}(x,u,w)$} denotes the energy cost \change{at prediction step $i$}. For example, \change{it} may represent thermal energy use multiplied by a \change{time-varying} energy price, or the electrical energy use of a heat pump whose coefficient of performance depends on outdoor temperature and supply-water temperatures~\cite{zanetti2023performance}.

In~\eqref{eq:nominal_mpc_slack}, the slack variable $s_{i|t}$ preserves feasibility when the predicted temperature cannot satisfy the nominal bounds exactly. In~\eqref{eq:nominal_mpc_obj}, the quadratic cost $\rho_\mathrm{slack} \, s_{i|t}^2$ penalizes this relaxation. 
However, tuning $\rho_\mathrm{slack}$ only affects temperature-bound violation indirectly. 
The relation between $\rho_\mathrm{slack}$ and the realized \acs{kh} violation is generally nonlinear~\cite{zheng2024economic}. In addition, the same slack weight can lead to different \acs{kh} outcomes across different weather conditions, buildings, and \acs{hvac} systems. Thus, a slack weight does not \change{let the user directly specify} the allowed violation severity.
These limitations motivate the Kh-budget MPC, which adds \acs{kh}-budget-aware components to the nominal \acs{mpc}.

\subsection{\acs{kh}-budget \acs{mpc}}
\label{sec:method_budget}

In the proposed \acs{kh}-budget \acs{mpc} framework, the \change{user} pre-specifies a running-average \acs{kh} budget $\alpha>0$ in $\mathrm{Kh/step}$. \change{The budget $\alpha$ corresponds to an average temperature-bound violation magnitude of $\frac{\alpha}{dT}$.} For example, with $dT=0.25~\mathrm{h/step}$, $\alpha=0.005$, $0.02$, and $0.05~\mathrm{Kh/step}$ \change{correspond to average violation magnitudes of} $0.02$, $0.08$, and $0.20~\mathrm{K}$, respectively. In closed-loop operation, the same average \change{can result from violations with different magnitudes and durations over time.}

\begin{figure}[!ht]
  \centering
  \includegraphics[width=0.95\linewidth]{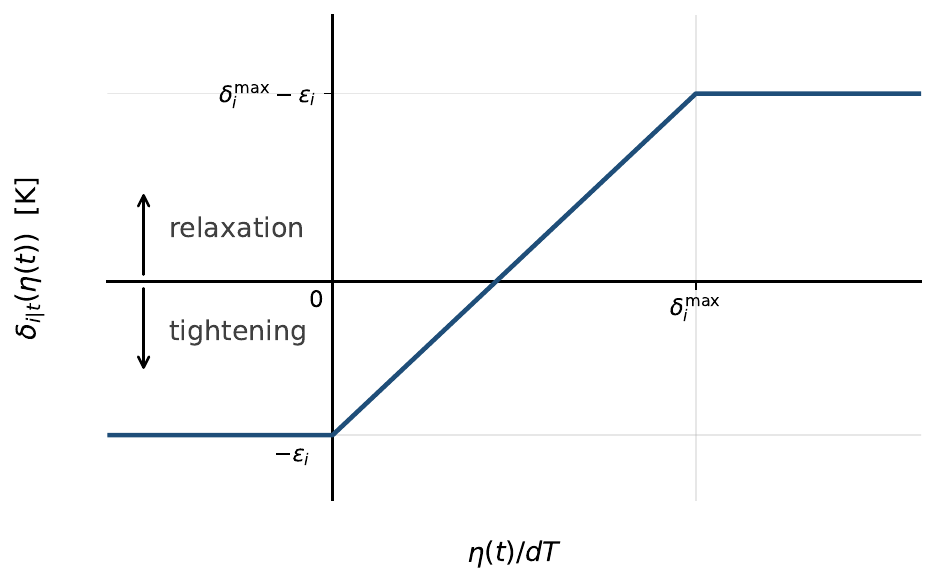}
\caption{\changeA{Illustration of $\delta_{i|t}(\eta(t))$.}}
  \label{fig:bound_adaptation_map}
\end{figure}

During operation, the controller uses this budget to update a budget-feedback state $\eta(t)$ and solve the \acs{kh}-budget \acs{mpc} problem. At time $t$, this state is updated after each realized violation measurement as
\begin{equation}
  \eta(t) = \eta(t-1) + K_P(\alpha-v(t)), \quad t\ge 1,
  \label{eq:eta_update}
\end{equation}
with $\eta(0)=0$. Then the following \acs{kh}-budget \acs{mpc} problem is solved:
\begingroup
\allowdisplaybreaks[4]
\begin{subequations}
\label{eq:budget_mpc}
\begin{align}
  \min_{u_{i|t},\,s_{i|t}}& \quad
   \sum_{i=0}^{N-1}
  l_{i|t}\!\left(x_{i|t},u_{i|t},w_{i|t}\right)
  + \sum_{i=1}^{N}\rho_\mathrm{slack} \, s_{i|t}^2
  \label{eq:budget_mpc_obj} \\
  \mathrm{s.t.}\quad
  & x_{0|t} = x(t), \\
  & x_{i+1|t} =
  f\!\left(x_{i|t},u_{i|t},w_{i|t}\right),  \ i=0,\ldots,N-1,\\
  & u_{i|t} \in \mathbb{U}, \quad i=0,\ldots,N-1,\\
  & y_{i|t} = g\!\left(x_{i|t}\right),
  \ i=1,\ldots,N, \\
& T^l_{i|t} - s_{i|t} - \delta_{i|t}(\eta(t))
\le y_{i|t} \label{eq:budget_mpc_bound}\\
& \quad \quad \le
T^u_{i|t} + s_{i|t} + \delta_{i|t}(\eta(t)),
\quad i=1,\ldots,N, \notag \\
  & s_{i|t} \ge 0,
  \ i=1,\ldots,N.
\end{align}
\end{subequations}
\endgroup
\changeA{The signed bound-adaptation signal in~\eqref{eq:budget_mpc_bound} is defined from the budget-feedback state $\eta(t)$ as}
\begin{equation}
\delta_{i|t}(\eta(t)) =
\begin{cases}
-\epsilon_i,
& \dfrac{\eta(t)}{dT}<0, \\[5pt]
\dfrac{\eta(t)}{dT}-\epsilon_i,
& 0\le\dfrac{\eta(t)}{dT}\le\delta_i^{\max}, \\[7pt]
\delta_i^{\max}-\epsilon_i,
& \dfrac{\eta(t)}{dT}>\delta_i^{\max},
\end{cases}
\quad i=1,\ldots,N.
\label{eq:budget_mpc_delta}
\end{equation}
\changeA{Figure~\ref{fig:bound_adaptation_map} illustrates this mapping: negative values tighten the temperature bounds, while positive values relax them.}
Here, $\epsilon_i$ and $\delta^{\max}_i$ are implementation parameters. Their roles and selection are discussed below. 
The full closed-loop procedure is summarized in Algorithm~\ref{alg:budget}.
\change{This feedback structure is related to running-average violation methods~\cite{korda2012stochastic,oldewurtel2013adaptively,shi2025dad}, and~\cite{oldewurtel2013adaptively} similarly adapts constraint tightening directly. However, these methods regulate violation rate rather than \acs{kh} severity and are formulated and evaluated on \acs{lti} systems.}

\begin{algorithm}[!ht]
\caption{Full procedure of the \acs{kh}-budget \acs{mpc} framework}
\label{alg:budget}
\begin{algorithmic}[1]
\STATE Choose the \acs{kh} budget $\alpha$, gain $K_P$, non-negative clipping values $\delta^{\max}_i$, and non-negative bound tightening values $\epsilon_i$. Set $\eta(0)=0$.
\FOR{$t=0,1,\ldots$}
  \STATE Observe the current state $x(t)$ and output $y(t)$.
  \IF{$t>0$}
    \STATE Measure the violation $v(t)$ from $y(t)$, $[T^l(t),T^u(t)]$, and $dT$ using equations~\eqref{eq:vtT}-\eqref{eq:vt}. Update the budget-feedback state $\eta(t)$ using~\eqref{eq:eta_update}.
  \ENDIF
    \STATE Solve the optimization problem in~\eqref{eq:budget_mpc}. Apply the first input $u(t)=u^\star_{0|t}$.
\ENDFOR
\end{algorithmic}
\end{algorithm}

In the following, Subsection~\ref{sec:method_budget_step} explains how the bound adaptation provides step-by-step budget feedback, and Subsection~\ref{sec:method_budget_detail} discusses implementation settings and the extension to multi-zone buildings.

\subsubsection{Bound adaptation and step-by-step budget feedback}\label{sec:method_budget_step}

The \acs{kh}-budget \acs{mpc} formulation in~\eqref{eq:budget_mpc} follows the nominal \acs{mpc} formulation~\eqref{eq:nominal_mpc} in its objective and most constraints. The key difference is the temperature constraint~\eqref{eq:budget_mpc_bound}. \changeA{It uses the signed bound-adaptation signal $\delta_{i|t}(\eta(t))$ to adapt the softened temperature bounds according to the budget-feedback state $\eta(t)$. Negative and positive values tighten and relax the bounds, respectively. At initialization, $\eta(0)=0$, so~\eqref{eq:budget_mpc_delta} gives $\delta_{i|t}(\eta(0))=-\epsilon_i$. Thus, the non-negative $\epsilon_i$ sets the initial tightening margin.}

Next, we explain how this bound adaptation provides step-by-step budget feedback. After the realized current-step violation $v(t)$ is measured, Algorithm~\ref{alg:budget} Step~5 updates the budget-feedback state as~\eqref{eq:eta_update}. The term $\alpha-v(t)$ is the current-step budget balance: $\eta(t)$ increases when the realized violation is below the per-step budget $\alpha$, and decreases when the realized violation exceeds it. 
Equation~\changeA{\eqref{eq:budget_mpc_delta}} maps $\eta(t)$ to \changeA{$\delta_{i|t}(\eta(t))$}. Thus, budget surplus \changeA{increases $\eta(t)$, increases relaxation,} and creates more energy-saving flexibility, while budget overspend \changeA{decreases $\eta(t)$, reduces relaxation,} and can make the \acs{mpc} more conservative.

Unrolling the $\eta(t)$ update clarifies how the per-step budget accumulates over time. With $\eta(0)=0$, recursively expanding~\eqref{eq:eta_update} gives
\begin{equation}
  \eta(t)
  =
  K_P\sum_{j=1}^{t}(\alpha-v(j))
  =
  K_P\left(\alpha t-\sum_{j=1}^{t}v(j)\right).
  \label{eq:eta_unrolled}
\end{equation}
Before the saturation in~\eqref{eq:budget_mpc_delta}, this update has the form of proportional feedback on the cumulative budget balance. Here, $\alpha t$ is the cumulative budget implied by the per-step budget $\alpha$, and $K_P$ is the proportional gain. 
As a result, when the running-average violation reaches or exceeds the budget, i.e., $(1/t)\sum_{j=1}^{t}v(j)\ge\alpha$, then $\eta(t)\leq 0$ and \changeA{\eqref{eq:budget_mpc_delta} gives $\delta_{i|t}(\eta(t))=-\epsilon_i$}. The optimized temperature bounds therefore retain the full tightening margin $\epsilon_i$. In this sense, $\alpha$ acts as the running-average reference that determines when \changeA{bound relaxation} is available.

\subsubsection{Implementation settings and multi-zone extension} \label{sec:method_budget_detail} 

\textbf{\textit{Prediction model and bound settings.}}
The \acs{kh}-budget formulation does not require a specific predictive model. The dynamics function $f(\cdot)$ in~\eqref{eq:f} may be data-driven, gray-box, or physics-based, provided that the tightening margins $\epsilon_i$ are chosen consistently with the selected predictor. 
These margins shape how the budget feedback modifies the temperature bounds, providing conservative operation when the accumulated budget surplus is small.
The margin can be selected from engineering experience~\cite{zheng2024economic} or from uncertainty-calibration techniques such as conformal prediction~\cite{angelopoulos2023conformal,shi2024disturbance} and the scenario approach~\cite{oldewurtel2013adaptively}, depending on the available information and the desired conservativeness. For example, conformal prediction has been used to account for horizon-step prediction errors in $\epsilon_i$, including model mismatch, measurement noise, and weather-forecast error~\cite{shi2024disturbance}.
The \changeA{clipping values $\delta_i^{\max}$ set the upper saturation in~\eqref{eq:budget_mpc_delta}}. A practical choice is
\[
\delta_i^{\max}=\epsilon_i+\bar{\delta}^{\max},
\]
\changeA{which gives}
\[
-\epsilon_i
\le
\delta_{i|t}(\eta(t))
\le
\bar{\delta}^{\max}.
\]
Here, $\bar{\delta}^{\max}\ge0$ is the user-specified maximum relaxation relative to the nominal temperature bounds. \changeA{Thus, separately from the softening variable $s_{i|t}$, the nominal bounds can be tightened by at most $\epsilon_i$ or relaxed by at most $\bar{\delta}^{\max}$.} For example, with nominal bounds of $21$--$25^\circ\mathrm{C}$ and $\bar{\delta}^{\max}=1~\mathrm{K}$, this component can relax the bounds to at most $20$--$26^\circ\mathrm{C}$.

\textbf{\textit{Multi-zone implementation.}}
The scalar formulation above is written for one controlled output. 
For a multi-zone building, the same \acs{kh}-budget feedback structure can be applied separately to each controlled zone, with zone-specific outputs, budgets, Kh violations, feedback states, and \changeA{bound-adaptation signals}. For zone $z$, the controller uses $\alpha_z$, $v_z(t)$,  $\eta_z(t)$, $\epsilon_{z,i}$ and \changeA{$\delta_{z,i|t}(\eta_z(t))$} to update that zone's temperature bounds. Coupling between zones, such as a shared heat pump, a shared supply-water temperature, or interacting thermal dynamics, is handled inside the \acs{mpc} model, constraints, and objective, while each zone keeps its own \acs{kh} budget.

\section{Case studies and validation configurations}
\label{sec:case_studies}

This section introduces the three building cases and describes the simulation and field configurations used to evaluate the \acs{kh}-budget \acs{mpc} framework.

\subsection{Validation logic and building cases}
\label{sec:validation_cases}

The three cases are selected to provide complementary validation evidence using five configurations, V1--V5, for \acs{kh}-budget MPC. Case~1 is a one-zone \acs{boptest} simulation benchmark~\cite{blum2021building} used for V1 Kh-budget response, V2 predictor portability, and V3 disturbance sensitivity. Case~2 is a two-zone \acs{boptest} simulation benchmark used for V4 coupled two-zone budget operation with symmetric and asymmetric zone budgets. Case~3 is an occupied residential deployment used for V5 field evaluation under real sensing, actuation, weather forecasts, and occupancy.

Case~1 is the BESTEST hydronic \ac{hp} test case from the \acs{boptest} framework. It represents a residential dwelling for a family of five, modeled as a single thermal zone in Brussels, Belgium, and served by an air-to-water modulating \ac{hp} with hydronic floor heating. The control-relevant occupancy schedule assumes occupied periods before 07:00 and after 20:00 on weekdays and throughout weekends. The temperature bounds are $21^\circ\mathrm{C}\le y\le25^\circ\mathrm{C}$ during occupied periods and $16^\circ\mathrm{C}\le y\le30^\circ\mathrm{C}$ during unoccupied periods.

Case~2 is the \acs{boptest} \texttt{twozone\_apartment\_hydronic} test case, representing a two-room residential apartment with one bathroom in Milan, Italy. The model has two controlled thermal zones. Each zone is served by an underfloor radiant loop with one on/off valve, and heat is supplied by a shared air-source \ac{hp}. In the reported simulations, the apartment is occupied by two people, one in each thermal zone, from 20:00 to 08:00 on weekdays and over the weekend. The same temperature bounds are used for both zones: $21^\circ\mathrm{C}\le y\le25^\circ\mathrm{C}$ during occupied periods and $16^\circ\mathrm{C}\le y\le30^\circ\mathrm{C}$ during unoccupied periods.

\begin{figure*}[!ht]
  \centering
  \begin{subfigure}[t]{0.49\textwidth}
    \centering
    \includegraphics[width=\linewidth]{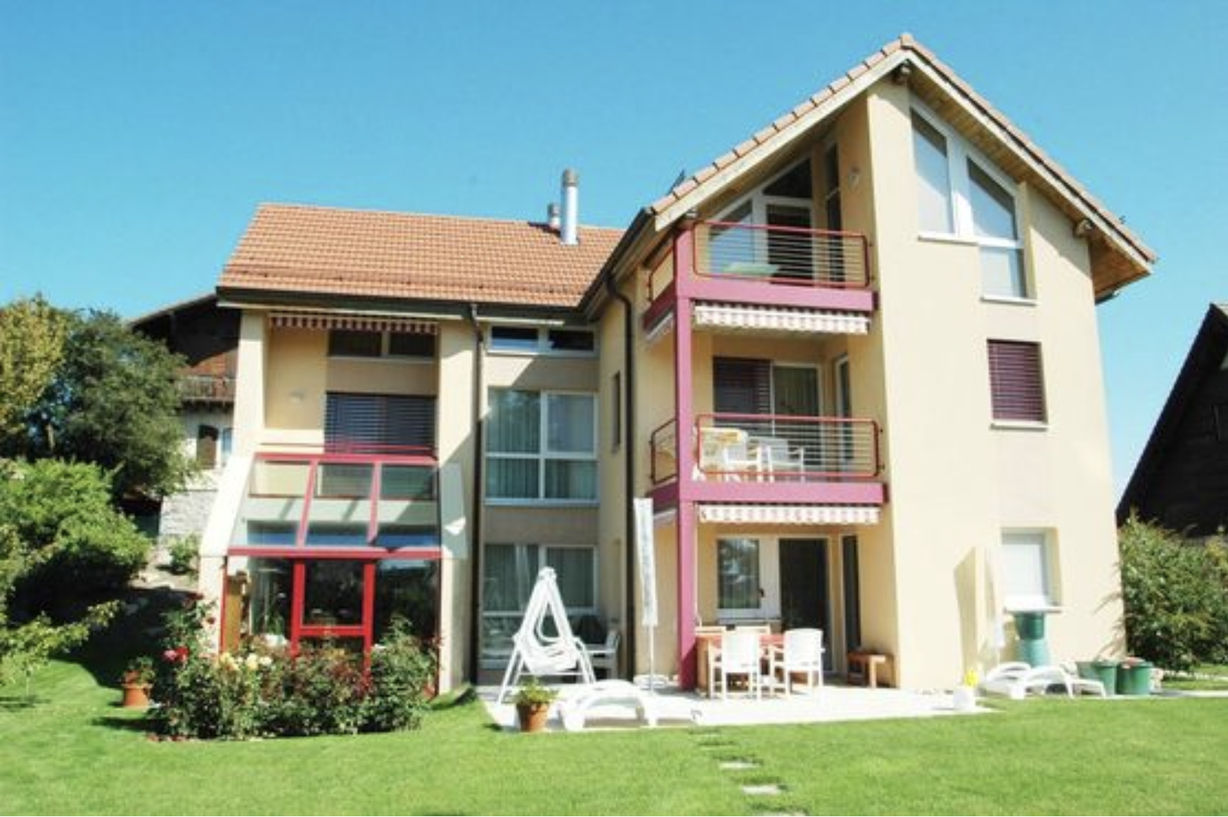}
    \caption{Occupied residential house}
    \label{fig:V5_house_site}
  \end{subfigure}
  \hfill
  \begin{subfigure}[t]{0.49\textwidth}
    \centering
    \includegraphics[width=\linewidth]{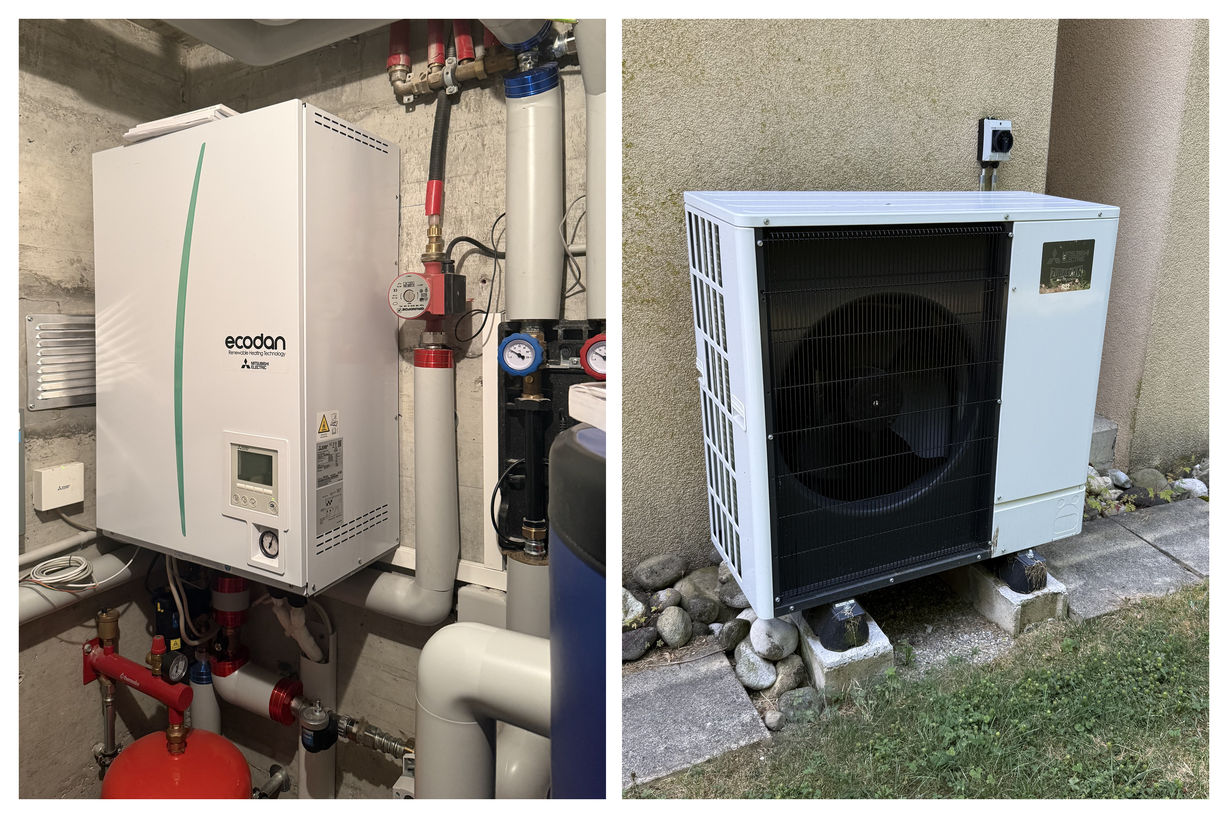}
    \caption{Heat-pump system}
    \label{fig:V5_house_hp}
  \end{subfigure}
  \caption{Occupied-home field site and Mitsubishi Ecodan heat-pump system.}
  \label{fig:V5_house}
\end{figure*}

Case~3 is an occupied single-family residential house in Switzerland, heated by a Mitsubishi Ecodan \ac{hp} with underfloor heating, as shown in Figure~\ref{fig:V5_house}. After discussions with the residents, one living-area floor was selected for monitoring and control, and it was treated as occupied throughout the experiment to avoid discomfort. The field comfort output is an ASHRAE~55-style operative temperature, $T_{\mathrm{op}}=\frac{26}{35}T_{\mathrm{room}}+\frac{9}{35}T_{\mathrm{floor}}$,
with coefficients calibrated from resident feedback. Following resident feedback, the lower temperature bound is reduced under higher solar \ac{ghi}: $T^l(t)=21.5^\circ\mathrm{C}$, $20.5^\circ\mathrm{C}$, and $19.5^\circ\mathrm{C}$ for $GHI<150$, $150\le GHI<300$, and $GHI\ge300~\mathrm{W/m^2}$, respectively. For Case~3, the \acs{kh} violation is calculated only for violations below the lower temperature bound.

\subsection{Simulation validation configurations}
\label{sec:case_sim_config}

The simulation evaluation includes four configurations, V1--V4. The shared simulation protocol and the case-specific controller implementation are summarized first.

\subsubsection{Simulation protocol and controller implementation}
\label{sec:sim_protocol}

All simulation runs use a sampling time of $15$ min. The prediction horizon is $N=48$, corresponding to $12$ h. The reported \acs{boptest} runs use the \texttt{peak\_heating\_day} scenario, i.e., a two-week test period centered on the day with the largest 15-min heating load of the year for the test case. The \acs{kh}-budget \acs{mpc} uses a slack weight $\rho_\mathrm{slack}=5$ and an update gain $K_P=0.25$ for $\eta(t)$. \change{The update gain $K_P=0.25$ was selected empirically and kept fixed across all \acs{kh}-budget \acs{mpc} simulation and field evaluations.} The maximum bound relaxation is set to $\bar{\delta}^{\max}=1.0~\mathrm{K}$ through $\delta^{\max}_i=\epsilon_i+\bar{\delta}^{\max}$. \change{Four disturbance settings $G_0$--$G_3$ combine two measurement-noise levels, $\sigma_n=0.1^\circ\mathrm{C}$ and $0.3^\circ\mathrm{C}$, with the \texttt{low} and \texttt{high} weather-forecast-error levels provided by the \acs{boptest} weather API. $G_0$ uses $\sigma_n=0.1^\circ\mathrm{C}$ and the \texttt{low} forecast-error level and serves as the common setting for V1, V2, and V4. V3 evaluates the other three combinations. The weather API uses the forecast-error emulator reported in~\cite{zheng2025quantifying}. Its parameter settings are given in~\ref{app:weather_error}.}

The common data-driven predictor used in V1, V3, and V4 is an adaptive \ac{arx} model with $n_a=n_b=6$, refitted hourly from recent data. The initial \acs{arx} model is fitted from 5 days of bang-bang-controller data in Case~1 and from 10 days of bang-bang-controller data in Case~2. The tightening margin $\epsilon_i$ is computed with \ac{scp} from the $98\%$ quantile of the absolute open-loop prediction residuals using a separate 5-day dataset~\cite{angelopoulos2023conformal,shi2024disturbance}.

For Case~1, the simulation controllers use room temperature as both the prediction state and controlled output, $x=y=T_{\mathrm{room}}$, and use the $15$-min average heat-pump electrical power as the manipulated input, $u=P_{\mathrm{elec}}$. The Case~1 \acs{mpc} objective is the sum of predicted heat-pump electrical energy. The \acs{mpc} returns the first-step optimized input, representing the average power target, for each $15$-min interval. It is implemented through a \acs{hp} on/off time allocation within each sampling period so that the realized average electrical power tracks the target. For Case~2, the shared \acs{hp}, shared supply-water temperature, zone valves, and fitted \acs{cop} model are described in~\ref{app:twozone_actuation_cop}.

All \acs{qp} realizations are formulated in \texttt{CVXPY}~\cite{diamond2016cvxpy} and solved with \texttt{PIQP}~\cite{schwan2023piqp}. V2 uses the same \acs{kh}-budget \change{\acs{mpc} formulation} with alternative predictor classes. A~\ac{nnarx} predictor used in V2 is formulated in \texttt{L4CasADi}~\cite{salzmann2024learning} and solved with \texttt{IPOPT}~\cite{wachter2006implementation}. V4 is formulated as a nonlinear program in \texttt{CasADi}~\cite{andersson2018casadi} and solved with \texttt{IPOPT}.

\change{Under the shared simulation setup described in this subsection, Table~\ref{tab:validation_config} summarizes the main differences across V1--V4 in case, predictor, disturbance setting, and tested parameters.}
\begin{table*}[!t]
\centering
\caption{\change{Summary of simulation validation configurations V1--V4.}}
\label{tab:validation_config}
\setlength{\tabcolsep}{4pt}
\renewcommand{\arraystretch}{1.08}
\begin{tabularx}{\textwidth}
{@{}llXX>{\centering\arraybackslash}p{0.10\textwidth}X@{}}
\toprule
\change{Config.} & \change{Case} & \change{Main variation} & \change{Predictor setting} & \change{Disturbance} & \change{Tested setting} \\
\midrule
\change{V1} & \change{Case~1} & \change{\acs{kh}-budget response and controller comparison} & \change{Adaptive \acs{arx}} & \change{$G_0$} & \change{$\alpha$ sweep; nominal \acs{mpc} and DAD-DPC sweeps} \\
\change{V2} & \change{Case~1} & \change{Prediction model} & \change{Adaptive data-driven; fixed gray-box; fixed neural-network} & \change{$G_0$} & \change{$\alpha=0.02,0.05$} \\
\change{V3} & \change{Case~1} & \change{Disturbance setting} & \change{Adaptive \acs{arx}} & \change{$G_1$--$G_3$} & \change{$\alpha=0.02,0.05$} \\
\change{V4} & \change{Case~2} & \change{Zone-budget assignment} & \change{Adaptive \acs{arx}} & \change{$G_0$} & \change{Three zone-budget pairs} \\
\bottomrule
\end{tabularx}
\end{table*}

\subsubsection{V1: \acs{kh}-budget response}
\label{sec:v1_direct_budget}

V1 evaluates the parameter-to-response behavior of the proposed \acs{kh}-budget \acs{mpc}. The configuration uses Case~1 under $G_0$ with the adaptive \acs{arx} predictor. 
The prescribed budget $\alpha$ is swept to test whether the achieved running-average \acs{kh} violation responds to the specified budget under step-by-step budget feedback. 
For comparison, the nominal \acs{mpc} with soft constraints~\eqref{eq:nominal_mpc} is swept over slack weights, and violation-rate DAD-DPC~\cite{shi2024disturbance} is swept over violation-rate targets.

\textbf{\acs{kh}-budget \acs{mpc}:} The proposed controller in Algorithm~\ref{alg:budget} is run with
\[
  \alpha \in \{0.005,\,0.010,\,0.020,\, 0.030,\,0.040,\,0.050\}
  ~\mathrm{Kh/step}.
\]

\textbf{Nominal \acs{mpc}:} The nominal \acs{mpc} in~\eqref{eq:nominal_mpc} is run with
\[
  \rho_\mathrm{slack}\in\{0.8,\,1,\,2,\,5,\,25,\,250\}.
\]
This sweep tests indirect slack-weight tuning, where temperature-bound violation is adjusted through the penalty on bound relaxation rather than through a prescribed \acs{kh} budget.

\textbf{DAD-DPC:} The DAD-DPC comparator is included because it pre-specifies an average violation-rate target rather than a \acs{kh} violation budget. \change{It adapts the disturbance bound used by its predictive controller based on a binary comfort-violation indicator to regulate the running-average violation rate. Here, $\beta$ is the prescribed violation-rate target, i.e., the allowed fraction of time steps outside the temperature bounds.} The target is swept as
\[
  \beta\in\{0.1,\,0.2,\,0.3,\,0.4,\,0.5,\,0.6 \}.
\]

\subsubsection{V2: Predictor portability}
\label{sec:v2_predictor}

V2 evaluates the predictor portability of the \acs{kh}-budget response by changing the prediction model within the \acs{kh}-budget \acs{mpc}.
The configuration uses Case~1 under $G_0$, with $\alpha\in\{0.02,\,0.05\}~\mathrm{Kh/step}$ and the remaining setup following Section~\ref{sec:sim_protocol}. The three additional predictors are an adaptive \change{\acf{deepc} predictor~\cite{lian2023adaptive}}, a fixed $4\mathrm{R}3\mathrm{C}$ gray-box \change{\acs{rc} model~\cite{drgovna2020all}}, and a fixed \change{\acs{nnarx} model~\cite{afroz2018modeling}} trained offline in \texttt{PyTorch}. These predictors test portability across online-adaptive and fixed models, linear and nonlinear structures, and data-driven and gray-box formulations, while keeping the \acs{kh}-budget \acs{mpc} formulation unchanged. More modeling details are given in~\ref{app:predictor_residual}.

\subsubsection{V3: Disturbance sensitivity}
\label{sec:v3_disturbance}

V3 evaluates the disturbance sensitivity of the \acs{kh}-budget response under stronger measurement-noise and weather-forecast-error conditions. V3 uses Case~1 with the adaptive \acs{arx} predictor and replaces the baseline disturbance setting $G_0$ with three stronger disturbance groups. $G_1$ uses a larger measurement-noise standard deviation, $\sigma_n=0.3^\circ\mathrm{C}$, with \texttt{low} forecast error; $G_2$ uses $\sigma_n=0.1^\circ\mathrm{C}$ and \texttt{high} forecast error; $G_3$ uses $\sigma_n=0.3^\circ\mathrm{C}$ and \texttt{high} forecast error. For each disturbance group, the proposed \acs{kh}-budget \acs{mpc} is run with $\alpha\in\{0.02,\,0.05\}~\mathrm{Kh/step}$. The remaining setup follows Section~\ref{sec:sim_protocol}.

\subsubsection{V4: Coupled two-zone budget operation}
\label{sec:v4_twozone}

V4 evaluates whether separate prescribed \acs{kh} budgets can be assigned in a coupled two-zone hydronic system. The configuration uses Case~2 under $G_0$ with the adaptive \acs{arx} predictor, and the remaining setup follows Section~\ref{sec:sim_protocol}. Three zone-budget settings are tested:
\begin{equation*}
    (\alpha_1,\alpha_2)\in\{(0.02,\,0.02),\,(0.02,\,0.05),\,(0.05,\,0.02)\}~\mathrm{Kh/step}.
\end{equation*}
The two zones share one \acs{hp} and one supply-water temperature, so the controller must allocate heat delivery through the two zone valves while applying separate zone-level running-average budget feedback. The three settings include one equal-budget case and two asymmetric assignments, so that the tighter budget is placed once on each zone.

\begin{figure*}[!ht]
  \centering
  \includegraphics[width=\linewidth]{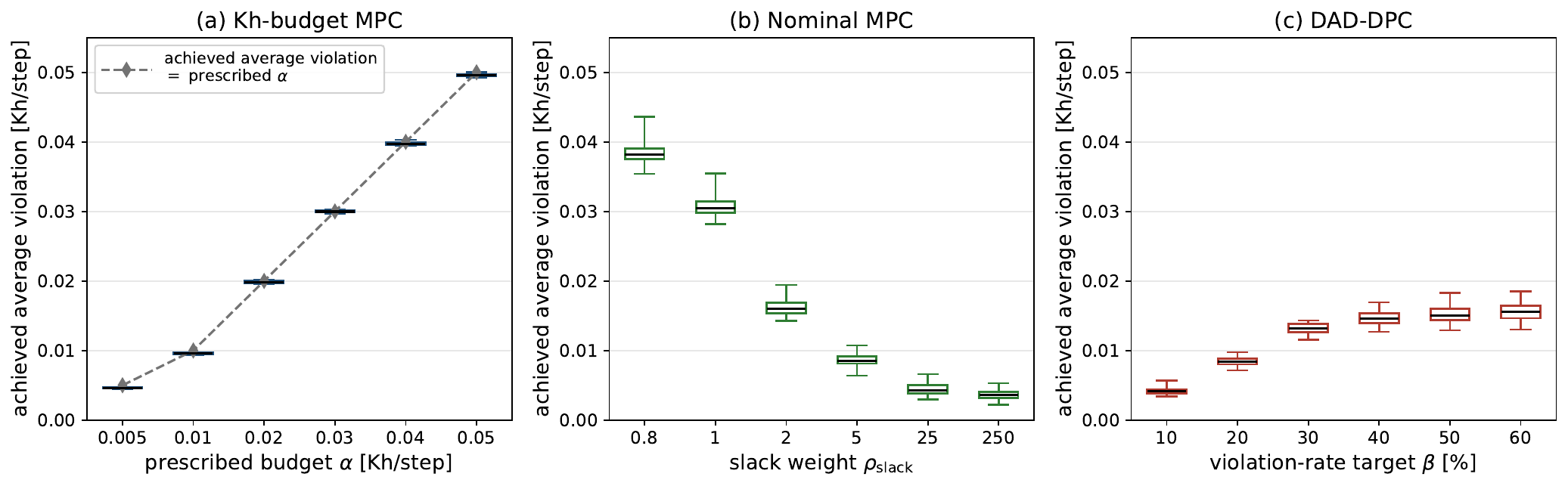}
  \caption{Monte Carlo distributions of achieved final average \acs{kh} violation in Case~1 for sweeps of the prescribed \acs{kh} budget $\alpha$, the nominal \acs{mpc} slack weight $\rho_\mathrm{slack}$, and the DAD-DPC violation-rate target $\beta$, respectively. In panel~(a), the dashed reference indicates achieved \acs{kh} violation equal to the prescribed $\alpha$.}
  \label{fig:v1_sweeps}
\end{figure*}

\subsection{Field deployment configuration}
\label{sec:case_field_config}

V5 evaluates occupied-home deployment of the \acs{kh}-budget \acs{mpc} in Case~3 under real sensing, actuation, weather, and occupancy conditions. The retrofit platform uses a Home Assistant Green hub running Home Assistant, Z-Wave air-temperature and floor-temperature sensors, and the control interface of the Mitsubishi Ecodan \ac{hp}.

The field deployment follows a retrofit supervisory architecture. The Home Assistant hub collects room-temperature and floor-temperature measurements from the Z-Wave sensors, and receives \ac{hp} measurements from the Mitsubishi \ac{hp} communication interface, including supply-water temperature and delivered heating power. An off-site computer at EPFL runs the \acs{mpc}, stores operation data in InfluxDB, receives field measurements from the Home Assistant hub, and sends supply-water-temperature setpoints back to the \ac{hp} through the same hub. Home Assistant retrieves weather forecasts from the Open-Meteo API at scheduled intervals.

V5 includes the default Weather Compensation mode of the Mitsubishi Ecodan \ac{hp}, the nominal \acs{mpc} in~\eqref{eq:nominal_mpc}, and the proposed \acs{kh}-budget \acs{mpc}. The Weather Compensation mode is normally used by the residents and serves as the default field controller. 
The nominal and \acs{kh}-budget \acs{mpc} controllers use an adaptive \acs{arx} predictor and optimize the delivered heating power $Q$.
This signal represents total underfloor-heating power and is used as a proxy measurement for heat delivered to the controlled zone, because the manually adjusted floor valves were kept fixed during the experiments.
Measured heat-pump electricity is not optimized, because the electrical meter also includes domestic hot-water production.
The predictor state is $x=[T_{\mathrm{room}},T_{\mathrm{floor}}]^\top$, the output is $y=T_{\mathrm{op}}$, the input is $u=Q$. \change{The field deployment uses the same \acs{scp} residual-calibration procedure as the simulation setup to compute the per-horizon-step tightening margin $\epsilon_i$ from the $98\%$ quantile of the absolute open-loop prediction residuals.} At the start of each episode, the initial adaptive \acs{arx} predictor with $n_a=n_b=6$ is fitted using the most recent $5$ days of field data, while the preceding $5$ days are \change{used for this residual calibration}. The predictor is then refitted hourly during operation. The prediction horizon is $N=32$, corresponding to $8$ h at the $15$-min sampling time. Both \acs{mpc}s are formulated as quadratic programming problems in \texttt{CVXPY} and solved with \texttt{PIQP}. The \acs{kh}-budget \acs{mpc} uses $\rho_\mathrm{slack}=5$, $K_P=0.25$, and  $\bar{\delta}^{\max}=0.5~\mathrm{K}$. Different parameters are tested: the nominal \acs{mpc} runs with $\rho_\mathrm{slack}\in\{50,\,5,\,1\}$ and the \acs{kh}-budget \acs{mpc} runs with $\alpha\in\{0.015,\,0.030,\,0.050\}~\mathrm{Kh/step}$.

During supervisory runs, the \ac{hp} is operated in a supply-water-temperature setpoint mode, and the retrofit actuation layer converts the \acs{mpc} heating-power target into minute-level supply-water-temperature setpoint updates without replacing the internal heat-pump controller.
The \acs{mpc} returns a $15$-min target heating power $Q^\star$ for the underfloor-heating loop. The supervisory controller updates the supply-water-temperature setpoint every minute, with bounded setpoint changes, so that the measured cumulative heating energy over the $15$-min cycle tracks $Q^\star dT$. 

\section{Simulation results}
\label{sec:results_sim}

We evaluate the proposed \acs{kh}-budget \acs{mpc} in simulation. V1 tests whether the prescribed \acs{kh} budgets produce clear running-average \acs{kh} responses and explains the associated bound adaptation. V2 and V3 \change{examine the \acs{kh}-budget response} across prediction models and disturbance groups. V4 tests zone-specific budgets in a thermally coupled two-zone setting. 

\subsection{Case~1: \acs{kh}-budget response and controller mechanism}
\label{sec:results_sim_response}

\begin{figure}[!ht] \centering \includegraphics[width=\linewidth]{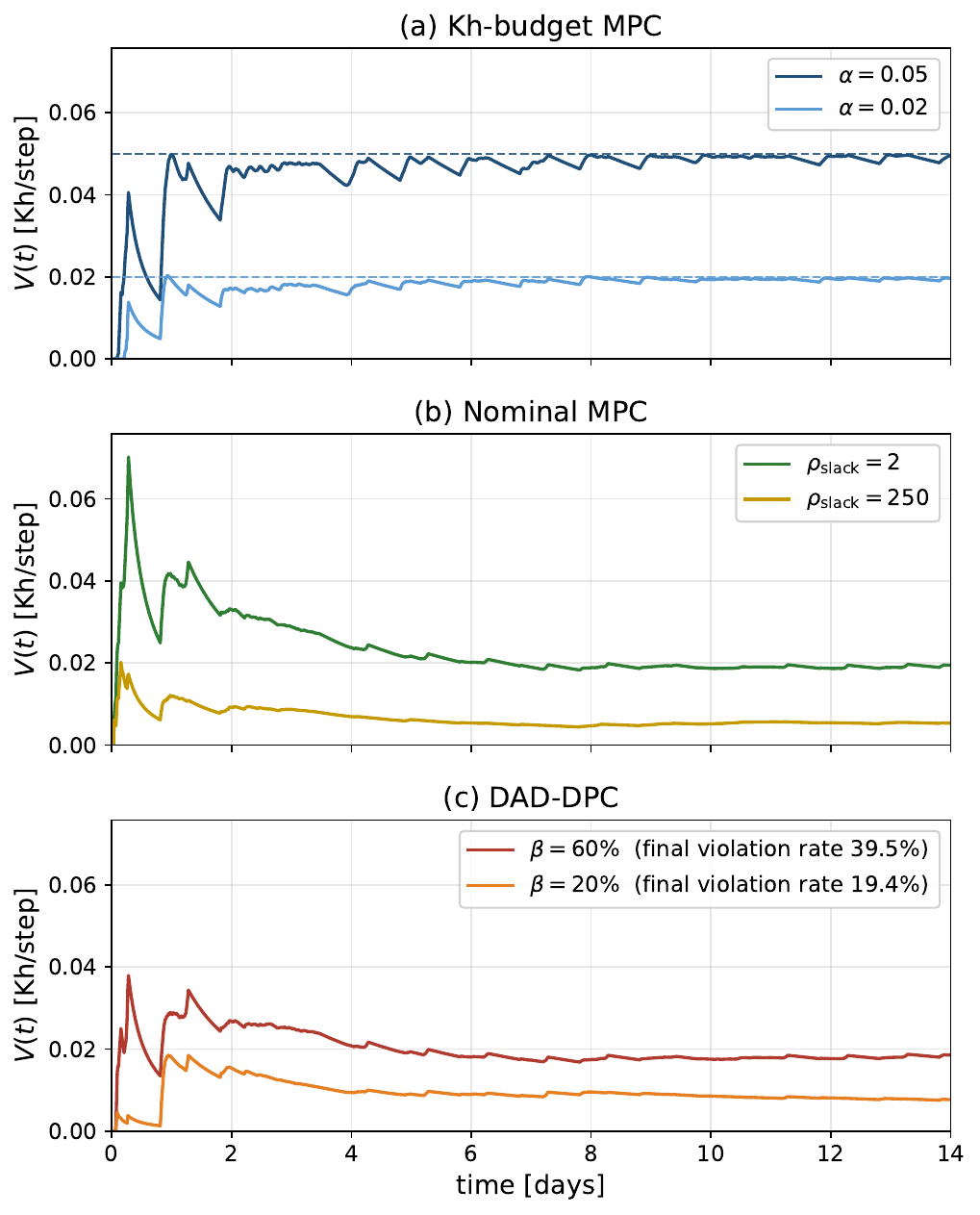} \caption{Running-average \acs{kh} violation $V(t)$ over the 14-day Case~1 horizon for one representative Monte Carlo realization. Dashed horizontal lines indicate the prescribed \acs{kh} budget for the proposed controller.} \label{fig:v1_running_average} \end{figure}

\begin{figure}[!ht] \centering \includegraphics[width=\linewidth]{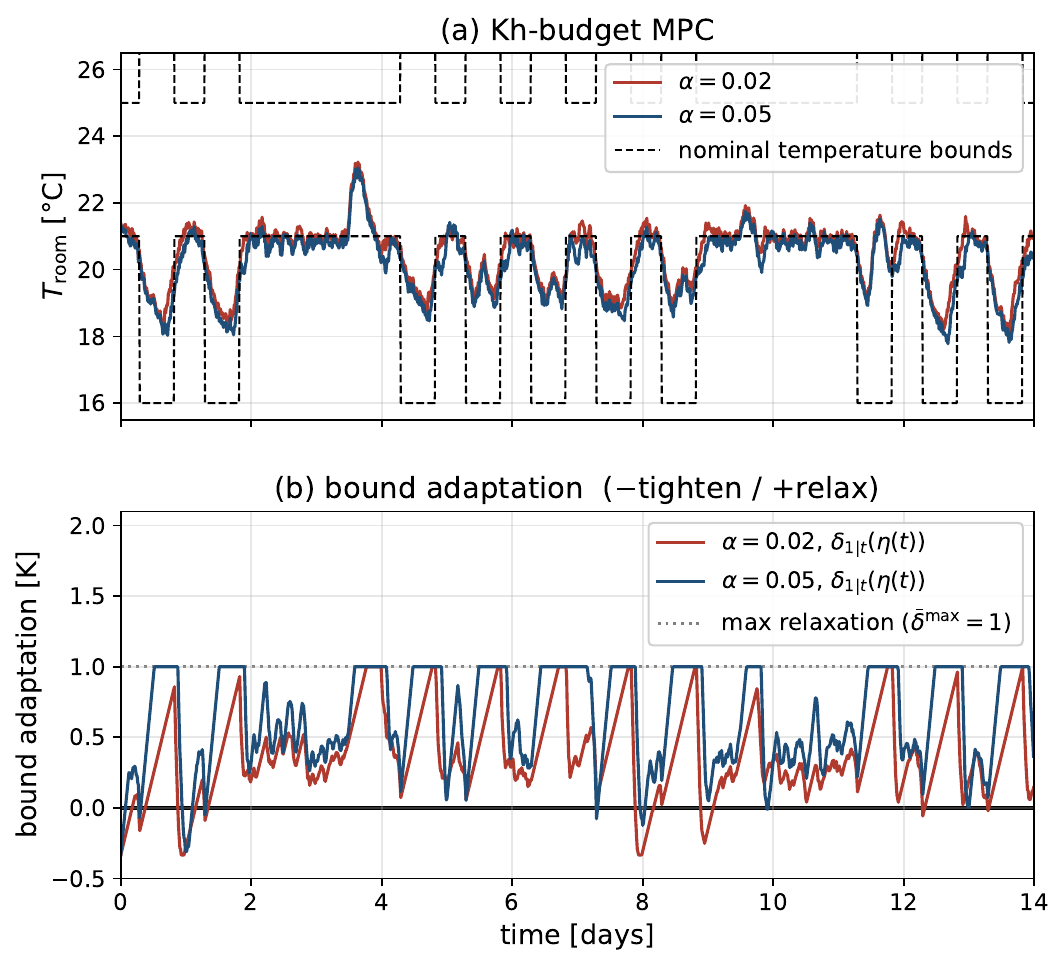} \caption{Representative operation and first-step bound adaptation of the proposed \acs{kh}-budget \acs{mpc} in Case~1. Positive values of $\delta_{1|t}(\eta(t))$ relax the temperature bounds.} \label{fig:v1_kh_mechanism} \end{figure}

\begin{figure}[!ht] \centering \includegraphics[width=\linewidth]{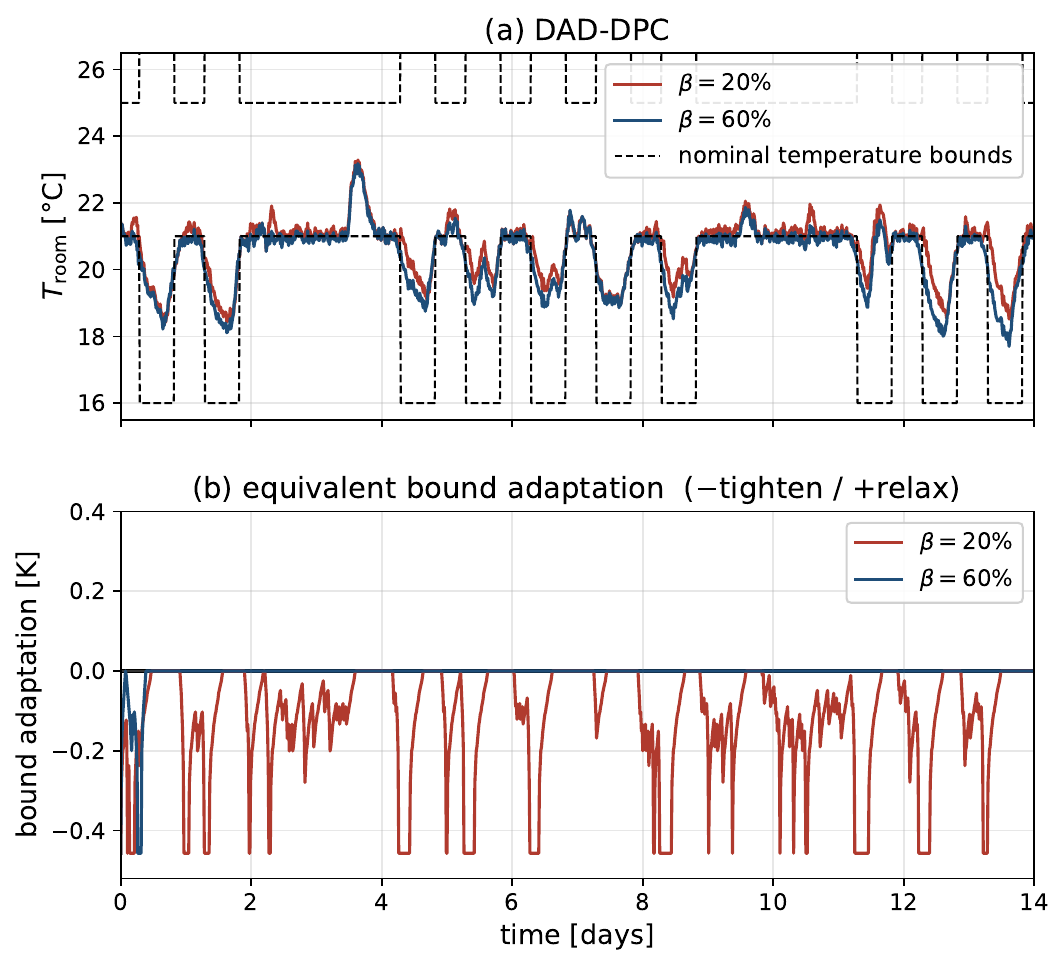} \caption{Representative operation of the DAD-DPC comparator in Case~1. The lower panel visualizes the adaptive disturbance-bound update as an equivalent temperature-bound adaptation.} \label{fig:v1_dad_mechanism} \end{figure}

Configuration~V1 compares the proposed \acs{kh}-budget \acs{mpc}, the nominal \acs{mpc}, and \ac{daddpc} on Case~1 under the same 20 Monte Carlo realizations of measurement noise and forecast error. Figure~\ref{fig:v1_sweeps} summarizes the parameter-to-outcome relation. Here, the final average \acs{kh} violation is the mean of the realized per-step \acs{kh} violations over the full run. For the proposed controller, this value stays close to the prescribed budget line, with small spread across runs. Nominal \acs{mpc} gives an indirect response: the achieved \acs{kh} violation changes nonlinearly with the slack weight $\rho_\mathrm{slack}$ and varies more across Monte Carlo runs. \acs{daddpc} targets violation rate rather than \acs{kh} severity, so $\beta$ cannot directly pre-specify a \acs{kh} budget. Its high-$\beta$ response also saturates in achieved \acs{kh} violation. Figures~\ref{fig:v1_running_average}--\ref{fig:v1_dad_mechanism} show the operation trajectories and adaptation signals behind the above outcomes.

Figure~\ref{fig:v1_running_average} shows the running-average \acs{kh} violation, defined as $V(t):=\frac{1}{t}\sum_{j=1}^{t} v(j)$, during closed-loop operation. For the proposed controller, $V(t)$ gets close to and remains mostly below the selected $\alpha$ values. This behavior shows that $\alpha$ is used online as a running-average budget. Nominal \acs{mpc} can reach a similar final \acs{kh} violation for some slack weights. For example, $\rho_\mathrm{slack}=2$ gives a final average violation of about $0.02$~Kh/step. Its running-average trajectory shows a large early transient, and the final value is obtained through indirect slack-weight tuning rather than by specifying a \acs{kh} budget. For DAD-DPC, the $60\%$ rate target gives a realized violation rate of only $39.5\%$, which is consistent with the limited increase in \acs{kh} violation observed in Figure~\ref{fig:v1_sweeps}. Periods with scheduled unoccupied daytime bounds can reduce $V(t)$ for all controllers because the lower temperature bound is reduced to $16~^\circ\mathrm{C}$ and violations are zero. 

Figure~\ref{fig:v1_kh_mechanism} shows the bound adaptation by the Kh-budget MPC controller and the room temperature trajectory during operation. When $V(t)$ is well below the prescribed budget, the \changeA{bound-adaptation signal $\delta_{1|t}(\eta(t))$ moves toward the positive relaxation limit}. As $V(t)$ approaches the budget, the relaxation is reduced. As a result, compared to $\alpha=0.02$, the looser budget $\alpha=0.05$ permits more relaxation and results in lower room temperature for energy reduction.

Compared with the Kh-budget MPC, DAD-DPC adapts the disturbance bound used in its \acs{mpc} to regulate a violation-rate target. Figure~\ref{fig:v1_dad_mechanism} visualizes this adaptation as an equivalent temperature-bound adaptation. The equivalent signal is non-positive, so DAD-DPC can reduce tightening toward zero but does not create positive relaxation beyond the nominal temperature bounds. This limits how much a high-violation-rate target can increase \acs{kh} severity, consistent with the saturation shown in Figure~\ref{fig:v1_sweeps}. 

\begin{figure}[!ht] \centering \includegraphics[width=\linewidth]{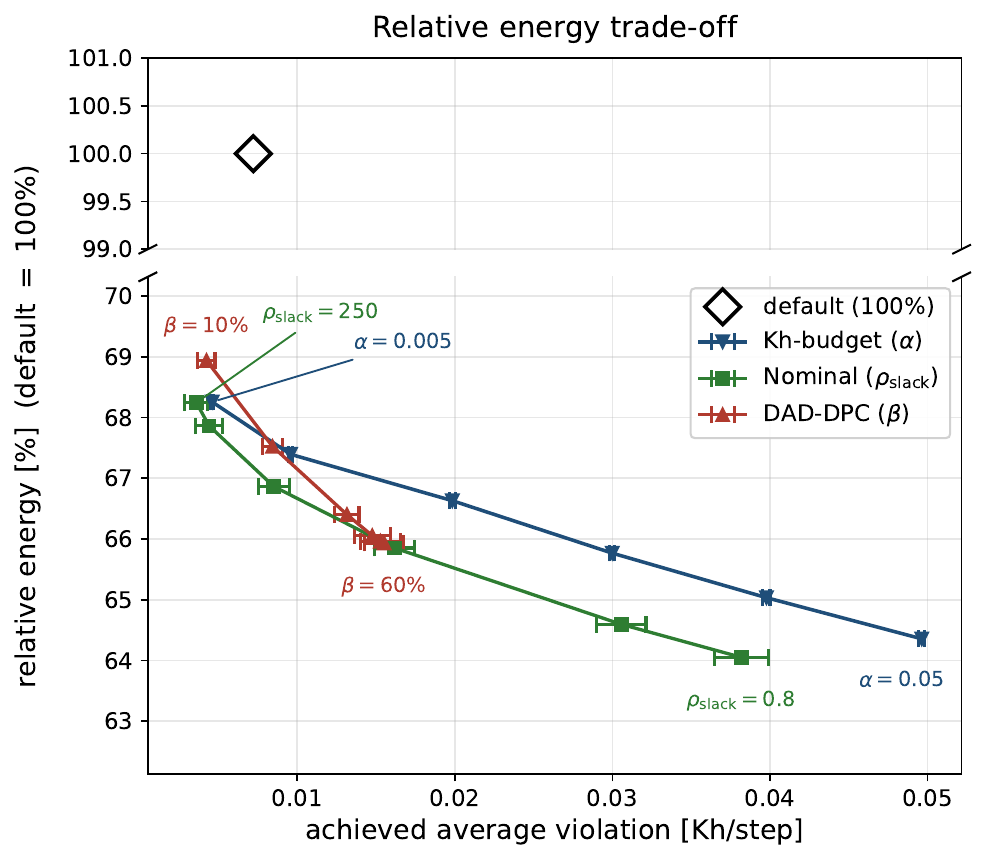} \caption{Relative energy use versus achieved \acs{kh} violation in Case~1. Controller markers correspond to sweeps of $\alpha$, $\rho_\mathrm{slack}$, and $\beta$, respectively. Energy use is normalized by the default controller, and horizontal error bars show one standard deviation of achieved \acs{kh} violation across the 20 Monte Carlo runs.} \label{fig:v1_tradeoff} \end{figure}

Figure~\ref{fig:v1_tradeoff} compares relative energy use and achieved \acs{kh} violation as the parameter of each controller family is swept. 
The default controller in Case~1 is the built-in \acs{boptest} PI controller on zone operative temperature, with occupied and unoccupied setpoints of $21.2^\circ\mathrm{C}$ and $20.5^\circ\mathrm{C}$, respectively.
All three controller families reduce energy use relative to the default controller over the tested parameter range, with average savings of about 31\% to 36\%. 
At the tested low-budget setting $\alpha=0.005~\mathrm{Kh/step}$, the proposed \acs{kh}-budget \acs{mpc} saves approximately $31.9\%$ energy while reducing achieved \acs{kh} violation by approximately $36.2\%$ relative to the default controller.
At a similar final average \acs{kh} violation, the proposed controller uses slightly more energy than nominal \acs{mpc}. This small difference is consistent with Figure~\ref{fig:v1_running_average}. Nominal \acs{mpc} reaches the final average after a large early transient, whereas the proposed controller spends more energy to respond to the prescribed running-average \acs{kh} budget during operation.

\subsection{Case~1: portability and disturbance checks}
\label{sec:results_sim_checks}

After establishing the parameter-to-outcome response of $\alpha$, we next \change{examine this relation} under predictor and disturbance changes. Configuration~V2 varies the prediction model while keeping the disturbance setting fixed. Configuration~V3 keeps the adaptive \acs{arx} predictor fixed and varies the measurement-noise and forecast-error levels.

Table~\ref{tab:v2_v3_checks} shows that the final average \acs{kh} violation is generally close to the prescribed budgets in all V2 predictor variants. The energy differences across predictors are small, suggesting that the \acs{kh}-budget MPC is not tied to a specific model class. 
In V3, the stronger disturbance groups retain a similar budget-response pattern: the average \acs{kh} violations scale with the prescribed budgets and remain close to them. The $G_3$ case at $\alpha=0.05~\mathrm{Kh/step}$ shows a weaker response and larger spread. This is consistent with the finite relaxation cap: once $\delta_{i|t}(\eta(t))$ reaches $\bar{\delta}^{\max}$, additional budget surplus cannot further relax the temperature bounds, so the looser budget is only partially expressed under the strongest disturbance setting.
Across the simulations at $\alpha=0.02$ and $0.05~\mathrm{Kh/step}$, the average energy use is reduced by $30.8\%$--$35.7\%$ relative to the default controller. 

\begin{table}[!ht]
\caption{Case~1 V2--V3 portability and uncertainty checks, with the default controller and the V1 adaptive \acs{arx} under $G_0$ included as references. Values are mean $\pm$ standard deviation across 20 Monte Carlo runs, and $\Delta E$ compares mean energy use with the default controller.}
\label{tab:v2_v3_checks}
\centering
\setlength{\tabcolsep}{3.5pt}
\renewcommand{\arraystretch}{1.08}
\begin{tabular*}{\columnwidth}{@{\extracolsep{\fill}}lccc}
\toprule
 &
\makecell{Achieved Kh\\($10^{-2}$ Kh/step)} &
\makecell{Energy\\(kWh/m$^2$)} &
\makecell{$\Delta E$ vs\\default} \\
\midrule
Default & $0.72 \pm 0.02$ & $3.48 \pm 0.00$ & -- \\
\midrule
\multicolumn{4}{@{}l}{Adaptive \acs{arx}, $G_0$ ($\sigma_n{=}0.1^\circ\mathrm{C}$, low forecast error)} \\
\quad $\alpha=$0.02 & $1.98 \pm 0.02$ & $2.32 \pm 0.01$ & $-33.4\%$ \\
\quad $\alpha=$0.05 & $4.96 \pm 0.02$ & $2.24 \pm 0.00$ & $-35.7\%$ \\
\midrule
\multicolumn{4}{@{}l}{\textit{V2 --- prediction model (with disturbance $G_0$)}} \\
\multicolumn{4}{@{}l}{Adaptive \acs{deepc}, $G_0$} \\
\quad $\alpha=$0.02 & $1.97 \pm 0.02$ & $2.33 \pm 0.01$ & $-33.1\%$ \\
\quad $\alpha=$0.05 & $4.93 \pm 0.04$ & $2.25 \pm 0.01$ & $-35.5\%$ \\
\addlinespace
\multicolumn{4}{@{}l}{\acs{rc}, $G_0$} \\
\quad $\alpha=$0.02 & $1.97 \pm 0.01$ & $2.31 \pm 0.01$ & $-33.6\%$ \\
\quad $\alpha=$0.05 & $5.00 \pm 0.02$ & $2.24 \pm 0.00$ & $-35.5\%$ \\
\addlinespace
\multicolumn{4}{@{}l}{\acs{nnarx}, $G_0$} \\
\quad $\alpha=$0.02 & $1.98 \pm 0.02$ & $2.34 \pm 0.03$ & $-32.8\%$ \\
\quad $\alpha=$0.05 & $5.02 \pm 0.03$ & $2.26 \pm 0.02$ & $-35.0\%$ \\
\midrule
\multicolumn{4}{@{}l}{\textit{V3 --- disturbance group (with adaptive \acs{arx})}} \\
\multicolumn{4}{@{}l}{$G_1$ ($\sigma_n{=}0.3^\circ\mathrm{C}$, low forecast error)} \\
\quad $\alpha=$0.02 & $2.00 \pm 0.03$ & $2.36 \pm 0.01$ & $-32.1\%$ \\
\quad $\alpha=$0.05 & $4.96 \pm 0.03$ & $2.26 \pm 0.01$ & $-35.0\%$ \\
\addlinespace
\multicolumn{4}{@{}l}{$G_2$ ($\sigma_n{=}0.1^\circ\mathrm{C}$, high forecast error)} \\
\quad $\alpha=$0.02 & $1.97 \pm 0.03$ & $2.35 \pm 0.02$ & $-32.3\%$ \\
\quad $\alpha=$0.05 & $4.89 \pm 0.06$ & $2.25 \pm 0.01$ & $-35.2\%$ \\
\addlinespace
\multicolumn{4}{@{}l}{$G_3$ ($\sigma_n{=}0.3^\circ\mathrm{C}$, high forecast error)} \\
\quad $\alpha=$0.02 & $1.94 \pm 0.06$ & $2.41 \pm 0.03$ & $-30.8\%$ \\
\quad $\alpha=$0.05 & $4.30 \pm 0.73$ & $2.30 \pm 0.03$ & $-33.9\%$ \\
\bottomrule
\end{tabular*}
\end{table}

\subsection{Case~2: coupled two-zone budget operation}
\label{sec:results_sim_twozone}

\begin{figure}[!ht]
  \centering
  \includegraphics[width=\linewidth]{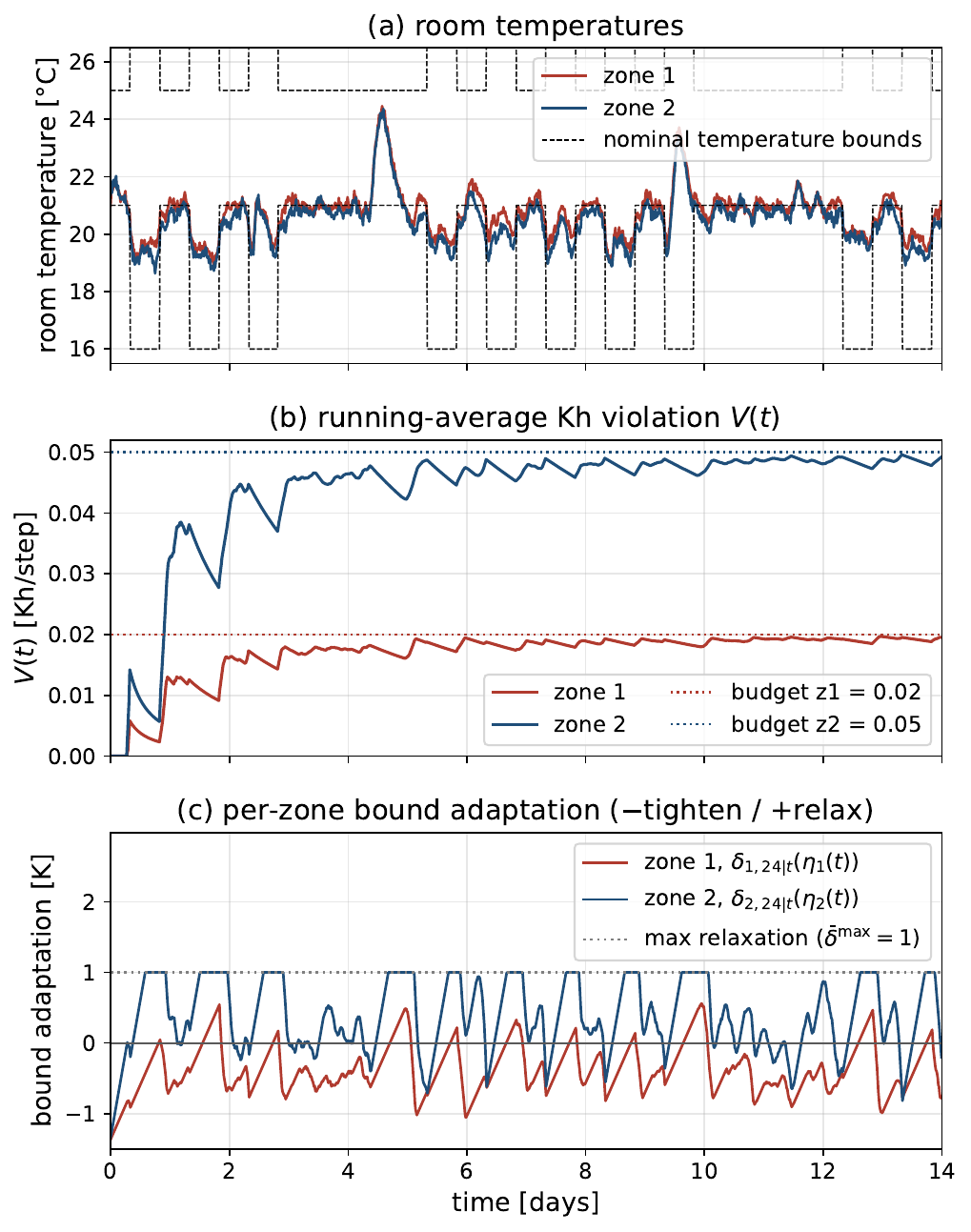}
  \caption{Representative two-zone operation for $(\alpha_1,\alpha_2)=(0.02,\,0.05)$. The lower panel shows the per-zone bound adaptation at the 24th prediction step.}
  \label{fig:v4_twozone}
\end{figure}

Configuration~V4 evaluates the per-zone response of the \acs{kh}-budget MPC in the coupled two-zone Case~2. Figure~\ref{fig:v4_twozone} shows the trajectories of one representative asymmetric setting, $(\alpha_1,\alpha_2) = (0.02,\,0.05)$. The running-average \acs{kh} trajectories reflect the asymmetric budgets: Zone~1 remains near the tighter budget, while Zone~2 realizes a higher \acs{kh} violation under the looser budget. The bound-adaptation signal is also zone-specific. In this case, the first-step bound-adaptation signals of both zones are often saturated at the relaxation limit. The lower panel therefore shows the values at the 24th step, where the asymmetric budget is more clearly reflected in the per-zone bound adaptation.

\begin{table}[!ht]
\caption{Case~2 V4 zone-specific \acs{kh}-budget results, with the default controller included as a reference. Values are mean $\pm$ standard deviation across 20 Monte Carlo runs. For non-default rows, the setting denotes $(\alpha_1,\alpha_2)$ in Kh/step.}
\label{tab:case2_two_zone}
\centering
\setlength{\tabcolsep}{3.5pt}
\renewcommand{\arraystretch}{1.08}
\begin{tabular*}{\columnwidth}{@{\extracolsep{\fill}}lccc}
\toprule
Setting &
\makecell{Zone 1\\($10^{-2}$ Kh/step)} &
\makecell{Zone 2\\($10^{-2}$ Kh/step)} &
\makecell{Energy\\(kWh/m$^2$)} \\
\midrule
Default
& $5.40 \pm 0.03$ & $5.43 \pm 0.03$ & $2.31 \pm 0.00$ \\
$(0.02,\,0.02)$
& $1.93 \pm 0.02$ & $1.94 \pm 0.02$ & $1.88 \pm 0.03$ \\
$(0.02,\,0.05)$
& $1.94 \pm 0.02$ & $4.91 \pm 0.04$ & $1.86 \pm 0.02$ \\
$(0.05,\,0.02)$
& $4.89 \pm 0.03$ & $1.94 \pm 0.02$ & $1.85 \pm 0.03$ \\
\bottomrule
\end{tabular*}
\end{table}

Table~\ref{tab:case2_two_zone} summarizes the Monte Carlo results for the symmetric and asymmetric budget settings. 
The default controller in Case~2 is the built-in \acs{boptest} rule-based controller: zone thermostats switch the floor-heating valves with a $2$~K hysteresis around the $21^\circ\mathrm{C}$ occupied and $16^\circ\mathrm{C}$ unoccupied setpoints, while the \ac{hp} supply-water temperature follows an outdoor-temperature climatic curve. 
The achieved \acs{kh} violations are reported in $10^{-2}$~Kh/step, while the $\alpha$ settings are given in Kh/step. Compared to the default controller, the tested \acs{kh}-budget settings reduce the mean energy across the 20 Monte Carlo runs by $18.9\%$--$20.5\%$. 
The zone-level mean \acs{kh} violation reductions are $9.4\%$--$64.3\%$, with smaller reductions in the looser-budget zones and larger reductions in the tighter-budget zones.
These results show that separate prescribed budgets produce zone-level responses in the thermally coupled two-zone case while reducing energy use relative to the default controller.

\section{Occupied-home field results} \label{sec:results_exp}

This section reports the occupied-home field results for Case~3. The field analysis examines \change{the running-average \ac{kh} response to the prescribed budgets} during online operation under real sensing, actuation, weather, and occupancy conditions. The deployment setup is summarized in Section~\ref{sec:case_field_config}.

\begin{table*}[!ht]
\centering
\caption{Occupied-home field run outcomes.}
\label{tab:field}
\small
\setlength{\tabcolsep}{3.5pt}
\begin{tabular}{l l l c c c c c c}
\toprule
Controller & Setting & Window & Duration & Final avg. & Heating/day & avg. $T_\mathrm{op}$ & avg. $T_\mathrm{out}$ & avg. GHI \\
& & (Year 2026) & (days) & Kh/step & (kWh/day) & ($^\circ$C) & ($^\circ$C) & (W/m$^2$) \\
\midrule
Default controller & --- & Feb. 20--26 & 6.0 & 0.0020 & 97.05 & 22.14 & 8.10 & 103 \\
Kh-budget MPC & $\alpha = 0.015$ & Mar. 22--28 & 6.0 & 0.0145 & 83.89 & 21.69 & 6.25 & 177 \\
Kh-budget MPC & $\alpha = 0.030$ & May 11--17 & 6.0 & 0.0278 & 62.56 & 21.64 & 9.65 & 214 \\
Kh-budget MPC & $\alpha = 0.050$ & Apr. 13--19 & 6.0 & 0.0284 & 27.31 & 21.66 & 12.63 & 225 \\
Nominal MPC & $\rho_{\mathrm{slack}} = 50$ & Mar. 28--30 & 2.0 & 0.0057 & 104.53 & 21.67 & 4.78 & 146 \\
Nominal MPC & $\rho_{\mathrm{slack}} = 5$ & Mar. 30--Apr. 1 & 2.0 & 0.0155 & 88.02 & 21.50 & 4.75 & 167 \\
Nominal MPC & $\rho_{\mathrm{slack}} = 1$ & Apr. 1--3 & 2.0 & 0.0537 & 67.29 & 21.34 & 6.55 & 228 \\
\bottomrule
\end{tabular}
\end{table*}

\begin{figure}[!ht] \centering \includegraphics[width=\linewidth]{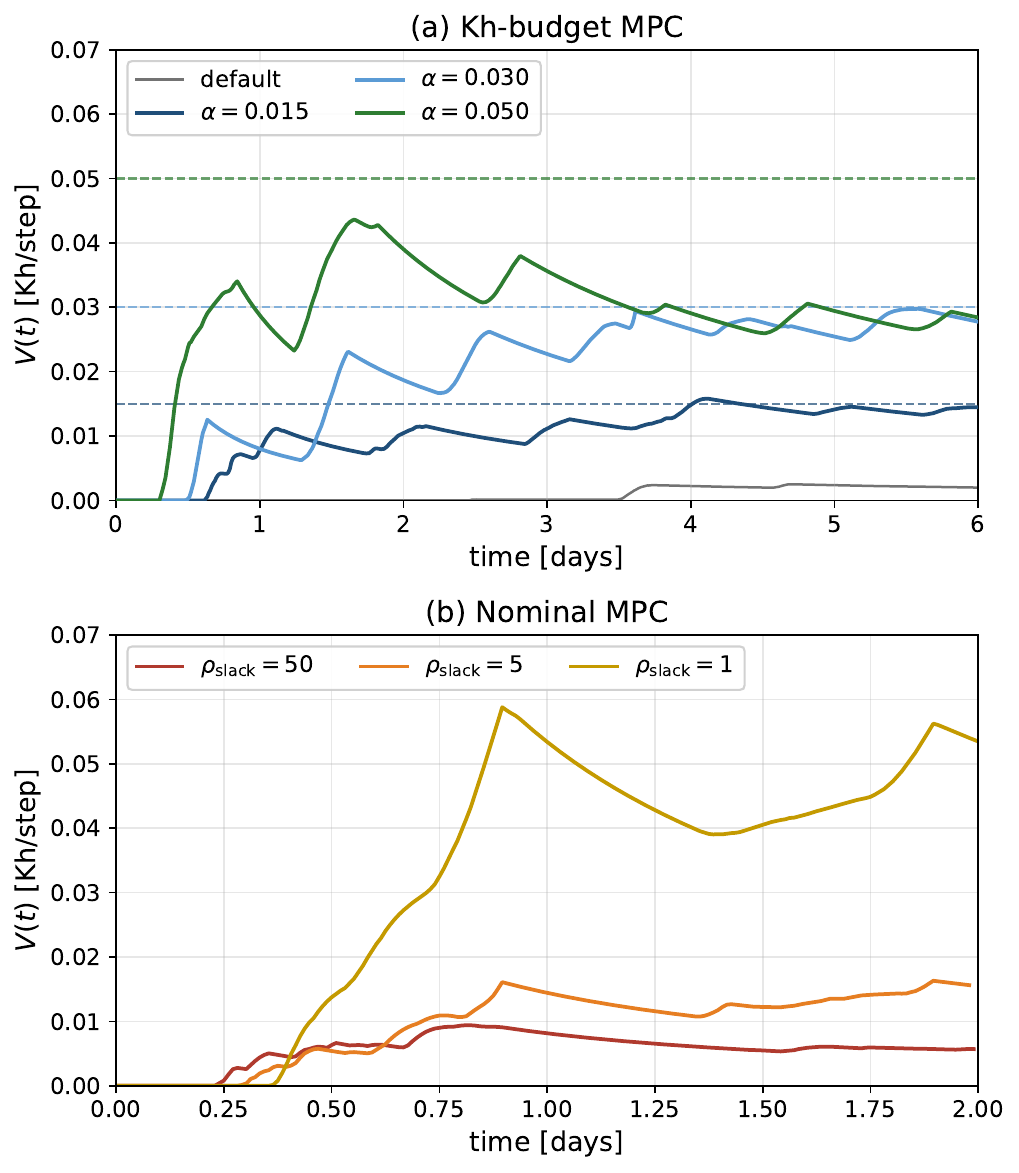} \caption{Field running-average \ac{kh} violation $V(t)$.} \label{fig:field_budget} \end{figure}

\begin{figure*}[!ht] \centering \includegraphics[width=\textwidth]{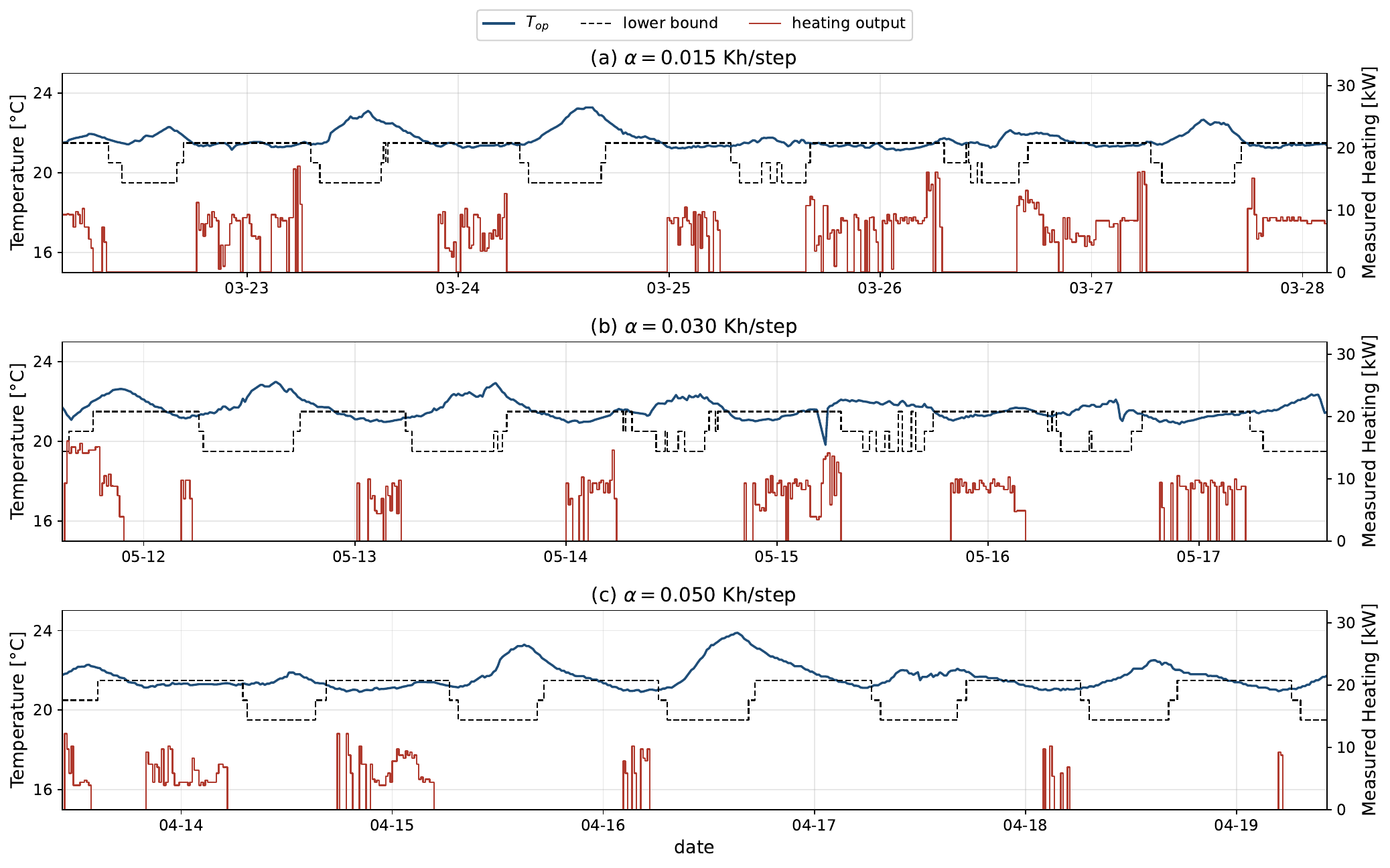} \caption{Field operative-temperature and measured heating-power trajectories for the \ac{kh}-budget \acs{mpc}. Panels (a)--(c) show $\alpha = 0.015$, $0.030$, and $0.050~\mathrm{Kh/step}$, respectively.} \label{fig:field_temp_heating_output} \end{figure*}

Table~\ref{tab:field} summarizes the occupied-home field runs. The \ac{kh}-budget \acs{mpc} completed all three six-day deployments. 
The $\alpha = 0.015$ and $0.030$ runs ended close to their prescribed budgets, with final average violations of $0.0145$ and $0.0278~\mathrm{Kh/step}$, respectively. The $\alpha = 0.050$ run ended well below its budget at $0.0284~\mathrm{Kh/step}$. This behavior is examined further using the closed-loop trajectories below.
The default controller was more conservative, ending near zero violation at $0.0020~\mathrm{Kh/step}$. Because the field runs span different outdoor temperatures and solar conditions, the heating-use values are reported as run-level observations with weather context, not as weather-matched energy comparisons.
The nominal \acs{mpc} runs produced final average violations from $0.0057$ to $0.0537~\mathrm{Kh/step}$ as $\rho_{\mathrm{slack}}$ decreased from $50$ to $1$. This tuning sensitivity shows that slack-weight tuning changes the realized violation, but it does not let the \change{user} prescribe a \ac{kh} budget before operation.

Figure~\ref{fig:field_budget} reports the running-average \ac{kh} violation trajectories. In Figure~\ref{fig:field_budget}(a), the $\alpha = 0.015$ and $0.030$ runs stay close to and mostly below their prescribed budgets after the initial transient. The $\alpha = 0.050$ trajectory remains well below its budget, indicating unused budget allowance during the mild late-spring window. The default controller stays close to zero throughout, reflecting a more conservative comfort outcome. Figure~\ref{fig:field_budget}(b) illustrates the contrast with the nominal \acs{mpc}: changing $\rho_{\mathrm{slack}}$ changes the running-average violation level, but the trajectories are not referenced to a prescribed \ac{kh} budget.

Figure~\ref{fig:field_temp_heating_output} shows the trajectories of the operative temperature and measured heating power for the three \ac{kh}-budget runs. 
The dashed lower-bound trajectories fluctuate during daytime as the GHI value changes, because the Case~3 setup applies a lower temperature bound during periods with higher GHI based on resident feedback, as described in Section~\ref{sec:validation_cases}.
The $\alpha = 0.015$ run keeps the operative temperature close to the lower bound and uses heating frequently. The $\alpha = 0.030$ run allows larger temperature deviations relative to the lower bound and ends close to its prescribed budget. The short temperature drop around the fourth day was caused by an occupant window-opening event.
In the $\alpha = 0.050$ run, measured heating drops to nearly zero during the final three days, but the mild conditions limit further \ac{kh} accumulation. This explains why the final average value remains well below the budget.

Taken together, the field results show that the \ac{kh}-budget \change{\acs{mpc}} can be deployed \change{in an occupied home and that the running-average \ac{kh} violations respond to the prescribed budgets under real operating conditions.}

\section{Discussion and conclusion}
\label{sec:conclusion}

This study proposes a \acs{kh}-budget \acs{mpc} for energy-efficient building control\change{, in which the user specifies a running-average \acs{kh} violation-severity budget before operation}. \change{The controller} uses this budget for step-by-step feedback during closed-loop operation. Signed bound adaptation adjusts the temperature bounds according to the realized \acs{kh} violation relative to the budget.
The validation covers high-fidelity simulation and occupied-home deployment. The simulations test budget response, predictor portability, disturbance sensitivity, and coupled two-zone operation. The field study tests whether the method remains applicable under real sensing, actuation, weather, and occupancy.

The simulation and field results show that prescribed \acs{kh} budgets produce clear running-average responses. In the one-zone simulation case, achieved \acs{kh} violations scale with the selected budgets and are generally close to or below them across predictors and disturbance settings.
At the low-budget setting $\alpha=0.005~\mathrm{Kh/step}$, the controller reduces energy use by $31.9\%$ and \acs{kh} violation by $36.2\%$ relative to the built-in default controller.
In the coupled two-zone simulation case, separate zone budgets produce zone-level responses, with $18.9\%$--$20.5\%$ energy reduction and $9.4\%$--$64.3\%$ zone-level \acs{kh} violation reduction. In the occupied-home deployment, the $\alpha=0.015$ and $0.030~\mathrm{Kh/step}$ runs ended close to their prescribed budgets, while the $\alpha=0.050~\mathrm{Kh/step}$ run ended well below its budget.

Several limitations remain. The field comparator runs were collected over different time windows and weather conditions. As a result, weather-matched quantitative energy comparisons cannot be made. Broader validation across buildings, climates, seasons, occupancy patterns, and disturbance conditions remains future work. 
The field temperature bounds were selected from occupant feedback and local measurements, including the \acs{ghi}-adaptive lower bound. Future work can combine the \acs{kh}-budget \change{framework} with occupant-centric comfort models that learn customized bounds. The same \change{approach} can also be tested in demand-response settings with time-varying prices or grid signals.

\appendix
\section{Implementation details for simulation configurations}
\label{app:implementation_details}

\subsection{Weather-forecast error emulator}
\label{app:weather_error}

The \texttt{low} and \texttt{high} weather-forecast-error levels used are generated by the \acs{boptest} weather API, which uses the forecast-error emulator reported in~\cite{zheng2025quantifying}. This appendix reports the API parameter settings. For outdoor temperature in $^\circ\mathrm{C}$,  the emulator uses
\begin{equation*}
e^T_{1|t}\sim\mathcal{N}(F_0,K_0^2),\qquad
e^T_{i+1|t}\sim\mathcal{N}(F e^T_{i|t}+\mu,K^2),    
\end{equation*}
where $i$ denotes the forecast step, and $\mathcal{N}$ denotes a normal distribution. The outdoor temperature settings are
\begin{equation*}
(F_0,K_0,F,K,\mu)=(0,0.6,0.92,0.4,0)    
\end{equation*}
for \texttt{low} and
\begin{equation*}
(F_0,K_0,F,K,\mu)=(-0.58,1.5,0.95,0.7,-0.015)   
\end{equation*}
for \texttt{high}. For \acs{ghi} in $\mathrm{W/m^2}$, the emulator uses
\begin{equation*}
e^G_{1|t}\sim\mathrm{Laplace}(a_{g,0},b_{g,0}),\qquad
e^G_{i+1|t}\sim\mathrm{Laplace}(\phi e^G_{i|t}+a_g,b_g),
\end{equation*}
where $\mathrm{Laplace}$ denotes a Laplace distribution.
The \ac{ghi} settings are
\begin{equation*}
(a_{g,0}, b_{g,0}, \phi, a_g, b_g)=(4.44,57.42,0.62,1.86,45.64)   
\end{equation*}
for \texttt{low} and
\begin{equation*}
(a_{g,0}, b_{g,0}, \phi, a_g, b_g)=(32.09,119.94,0.67,10.63,87.44)  
\end{equation*}
for \texttt{high}. For \acs{ghi}, the emulator also sets nighttime forecasts to zero and applies smoothing post-processing to nonzero forecasts. See~\cite{zheng2025quantifying} for more details.

\subsection{Two-zone implementation and \acs{cop}-aware objective for Case~2}
\label{app:twozone_actuation_cop}

For Case~2, the adaptive \acs{arx} predictor uses
\begin{align*}
x(t)&=
\begin{bmatrix}
T_{\mathrm{room},1}(t) & T_{\mathrm{room},2}(t) & T_{\mathrm{floor},1}(t) & T_{\mathrm{floor},2}(t)
\end{bmatrix}^{\top},\\
y(t)&=
\begin{bmatrix}
T_{\mathrm{room},1}(t) &
T_{\mathrm{room},2}(t)
\end{bmatrix}^{\top},
\end{align*}
where $T_{\mathrm{room},z}$ and $T_{\mathrm{floor},z}$ are the room and floor temperatures of zone~$z$. The predictor input for zone $z$ is
\begin{equation*}
u_z(t)=\left(T_{\mathrm{sup}}(t)-T_{\mathrm{floor},z}(t)\right)\lambda_z(t),
\, z\in\{1,2\},
\end{equation*}
where $T_{\mathrm{sup}}(t)$ is the shared supply-water temperature and $\lambda_z(t)\in[0,1]$ is the valve duty fraction. In the optimization implementation, equivalent zone-level virtual variables are used to avoid placing the product $T_{\mathrm{sup}}(t)\lambda_z(t)$ directly in the predictor constraints.

The electrical-energy objective uses a fitted \acs{cop} model. Since the loop flow rate is fixed when a valve is open, $u_z(t)$ is proportional to heat delivered to zone $z$, up to a water-flow and heat-capacity factor. This factor is omitted in the objective. The fitted model is
\begin{equation*}
\mathrm{COP}(t)=c_1+c_2T_{\mathrm{sup}}(t)+c_3T_{\mathrm{out}}(t),
\end{equation*}
where $T_{\mathrm{out}}(t)$ is the outdoor-temperature component of $w(t)$. The coefficients are fitted once from the same $10$-day bang-bang-control data used for the initial Case~2 \acs{arx} model and are kept fixed during the V4 episode. In the prediction horizon, the electrical-energy term optimized by the \acs{mpc} is
\begin{equation*}
\sum_{i=1}^{N}
\frac{u_{1,i|t}+u_{2,i|t}}
{c_1+c_2T_{\mathrm{sup},i|t}+c_3T_{\mathrm{out},i|t}}.
\end{equation*}

For the implementation of MPC's optimal input, the optimized supply water temperature is sent as the setpoint of supply water temperature in BOPTEST. The zone valves accept only on/off commands, so the on-duration of the valve in zone~$z$ is chosen such that the sampling-period average approximates the optimized duty fraction $\lambda_z$.

\subsection{Modeling details for V2}
\label{app:predictor_residual}

For V2 predictor portability in Case~1, the adaptive \acs{arx} predictor follows the setup in Section~\ref{sec:sim_protocol}. The \acs{deepc} predictor uses the bilevel adaptive \acs{deepc} formulation of~\cite{lian2023adaptive}. Its non-parametric predictor is built from Hankel matrices with an initial-step length $d_{\mathrm{init}}=6$ and is updated hourly by rebuilding the Hankel representation from a rolling $5$-day data buffer. The fixed $4\mathrm{R}3\mathrm{C}$ \acs{rc} predictor is a gray-box thermal model with indoor, envelope, and floor thermal capacitances~\cite{drgovna2020all}. It is fitted offline using $30$ days of bang-bang-controller data. The fixed \acs{nnarx} predictor~\cite{afroz2018modeling} uses a four-step input-output history window and a feedforward network with three hidden layers, five neurons per layer, hyperbolic-tangent activations, and a residual connection. It is trained offline in \texttt{PyTorch} using 30 days of bang-bang-controller data and 30 days of \ac{prbs} data.

All four V2 predictors use the same \acs{scp} residual calibration protocol to compute the tightening margins. For each predictor, open-loop $N$-step residuals are computed with forecast weather on a held-out $5$-day dataset. The $98\%$ quantile of the absolute residuals is then computed at each horizon step and kept fixed during each episode.

\section*{CRediT authorship contribution statement}
\textbf{Jicheng Shi:} Conceptualization, Methodology, Software, Formal analysis, Investigation, Data curation,  Writing - original draft, Visualization.
\textbf{Colin N. Jones:} Conceptualization, Software, Validation, Writing - Review \& Editing, Resources, Supervision, Funding acquisition, Project administration.



\section*{Declaration of Generative AI and AI-assisted technologies in the writing process}
During the preparation of this \change{study}, the authors used ChatGPT in order to improve the readability and language and to check spelling and grammar. After using this tool, the authors reviewed and edited the content as needed and take full responsibility for the content of the publication.

\section*{Acknowledgement}
This \change{study} was supported by the Swiss National Science Foundation (SNSF) under the NCCR Automation project, grant agreement 51NF40 180545.



 \bibliographystyle{elsarticle-num} 
 \bibliography{refs_doi}

\end{document}